\documentclass[nonacm,sigplan]{acmart}
\usepackage{amsmath,amsfonts}
\usepackage{graphicx}
\usepackage{textcomp}
\usepackage{xcolor}
\usepackage{fancyhdr}
\usepackage[]{hyperref}
\usepackage{subcaption}
\usepackage{algorithm}
\usepackage{algpseudocode}
\usepackage{braket}
\usepackage{booktabs}  

\begin{document}

\title{\huge A Case for Quantum Circuit Cutting for NISQ Applications: \\{\LARGE Impact of topology, determinism, and sparsity}}

\author{Zirui Li}
\email{zirui.li@rutgers.edu}
\affiliation{%
  \institution{Rutgers University}
  \country{USA}
}
\author{Minghao Guo}
\email{mg1998@scarletmail.rutgers.edu}
\affiliation{%
  \institution{Rutgers University}
  \country{USA}
}
\author{Mayank Barad}
\email{mb1961@scarletmail.rutgers.edu}
\affiliation{%
  \institution{Rutgers University}
  \country{USA}
}
\author{Wei Tang}
\email{weit@alumni.princeton.edu}
\affiliation{%
  \institution{Princeton University}
  \country{USA}
}
\author{Eddy Z. Zhang}
\email{eddy.zhengzhang@gmail.com}
\affiliation{%
  \institution{Rutgers University}
  \country{USA}
}
\author{Yipeng Huang}
\email{yipeng.huang@rutgers.edu}
\affiliation{%
  \institution{Rutgers University}
  \country{USA}
}

\begin{abstract}

We make the case that variational algorithm ansatzes for near-term quantum computing are well-suited for the quantum circuit cutting strategy.
Previous demonstrations of circuit cutting focused on the exponential execution and post-processing costs due to the cuts needed to partition a circuit topology, leading to overly pessimistic evaluations of the approach.
This work observes that the ansatz Clifford structure and variational parameter pruning significantly reduce these costs.
By keeping track of the limited set of correct subcircuit initializations and measurements, we reduce the number of experiments needed by up to $16\times$, matching and beating the error mitigation offered by classical shadows tomography.
By performing reconstruction as a sparse tensor contraction, we scale the feasible ansatzes to over 200 qubits with six ansatz layers, beyond the capability of prior work.


\end{abstract}

\maketitle 
\pagestyle{plain} 

\section{Introduction}

There is significant excitement about when quantum computing might first become useful~\cite{Assessing_the_quantum-computing_landscape}.
The first applications of QC are expected to be in the simulation of quantum systems~\cite{Disentangling_Hype}.
Furthermore, it is expected that any early applications of QC will require tight cooperation between classical and quantum computing.
Although proposals of hybrid quantum-classical computing seem pragmatic, exactly how the two computing paradigms might interact is an open question.
The ways in which problems, algorithms, and solution data are represented in QC and classical computing are vastly different and do not allow for tight coupling~\cite{mike_and_ike,Kaye,mermin_2007}.

One straightforward strategy is to phrase problems and algorithms entirely in the QC paradigm and allow classical computing to augment the capability of QCs~\cite{Large_Small}.
In such an approach, quantum programs, known as quantum circuits, are carefully split apart such that the program fragments run independently.
The fragments are expected to perform with greater success in QCs due to their lower hardware requirements, shorter execution duration, and easier compilation of the problems in terms of hardware constraints.
A classical computer gathers the data collected from the individual QC experiments and reconstructs the complete result~\cite{CutQC,tang2022scaleqcscalableframeworkhybrid,kan2024scalablecircuitcuttingscheduling,Gate_Cuts_and_Wire_Cuts,pawar2023integratedqubitreusecircuit,Clifford-based_Circuit_Cutting}.
The interaction of classical and quantum hardware and software leads to architectural questions.

The strategy has significant momentum.
Various research studies have investigated the most beneficial way to decompose circuits to fit in constraints~\cite{CutQC,tang2022scaleqcscalableframeworkhybrid,kan2024scalablecircuitcuttingscheduling,Gate_Cuts_and_Wire_Cuts,pawar2023integratedqubitreusecircuit}, along with the most efficient way to perform experiment measurements that collect data for classical reconstruction~\cite{Golden_Circuit_Cutting_Points,Classical_Shadows,Approximate_Reconstruction}.
Industrial strength libraries for planning reconstruction computation~\cite{Gray2021hyperoptimized} and for performing actual reconstruction on GPUs~\cite{bayraktar2023cuquantumsdkhighperformancelibrary} all highlight the expectation that the strategy will have a practical impact.

The strategy also has critics.
It is understood that the number of cuts that are used to split the quantum programs---the \emph{topology} of a quantum circuit and its partitioning~\cite{Treewidth,Markov_tensor}---dictates the both the cost of subcircuit executions and the cost of reconstruction classical postprocessing.

This paper considers these costs by exploiting orthogonal and previously overlooked opportunities.

First, in quantum algorithms where a quantum circuit is run repeatedly with different parameters~\cite{Peruzzo2014,farhi2014quantum}, a lot of information can be gathered once the circuit topology and cutting plan are established.
In fact, the states to which subcircuits must be initialized and measured are already known \emph{deterministically} at this point.
The challenge in exploiting this knowledge is how to collect, compile, and query this information efficiently to drive the circuit cutting strategy.
In this work, we address this challenge using established techniques in classical artificial intelligence to reason about deterministic information in otherwise probabilistic models~\cite{kc_map}.
We apply these techniques to efficiently perform circuit cutting experiments with minimal sets of initializations and measurements, thus reducing the number of experiments needed to reach a given level of precision by $16\times$.
The compiled knowledge furthermore creates an opportunity to reduce error in the reconstructed result, thus matching or beating existing error mitigation approaches in quantum state tomography.

Second, because of the structure and determinism in the data collected from the quantum subcircuit executions, the measurement data are extremely \emph{sparse}.
Specifically, of the possible initializations and measurement outcomes that are possible, only very few sets have coefficients that contribute to the reconstructed state.
In practice, this means that previous approaches to performing reconstruction as conventional tensor multiplication have had significantly inflated and conservative cost estimates.
For nontrival circuits, we demonstrate a reduction in memory required by more than $90\%$, and we expand the size of ansatzes that are feasible in the circuit strategy to over 200 qubits, surpassing prior work.


\section{Motivation \& Background}

Current quantum computer prototypes are limited in both capacity and accuracy.
Researchers believe that the first applications of noisy intermediate-scale quantum (NISQ) computing would be through variational quantum algorithms (VQAs)~\cite{Preskill2018quantumcomputingin,national2019quantum}.
This section reviews existing work on VQAs, focusing on how they are especially suited for the quantum circuit cutting strategy.
Then, we review the necessary mathematical background for VQAs and circuit cutting.

\begin{figure}[t]
    \centering
    \includegraphics[width=\linewidth]{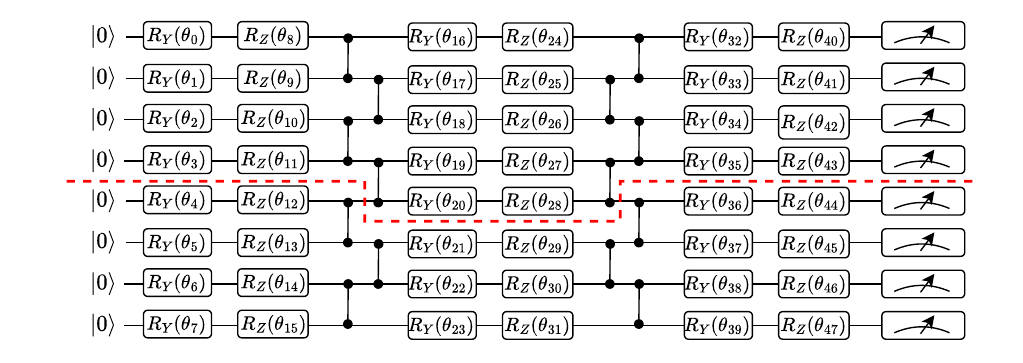}
    \caption{
The circuit cutting plan for HWEA with \#qubits 8, depth 2.
The qubit constraint for each subcircuit is 5.
We use HWEA (8,2,5) to describe this circuit cutting workload.
}
    \label{fig:HWEA_cut}
\end{figure}

\begin{figure}[t]
        \centering
        \includegraphics[width=\linewidth]{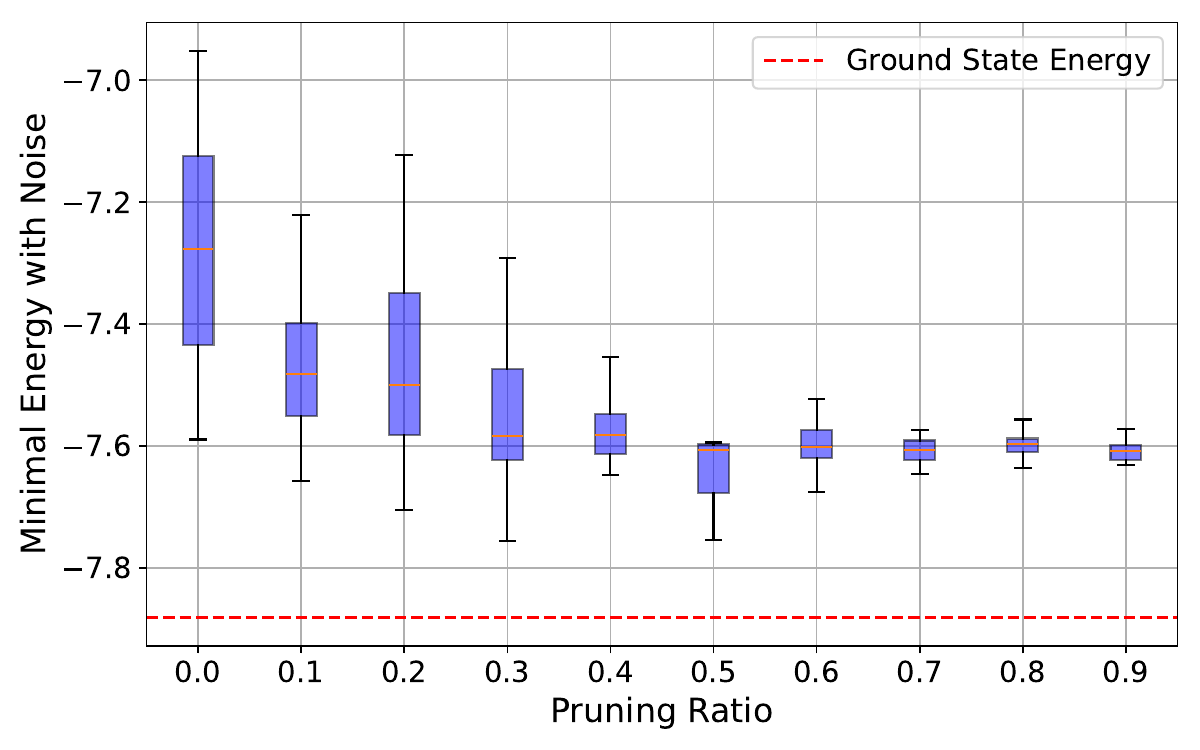}
        \caption{
Pruning non-Clifford gates with small rotation angles is an effective strategy for NISQ ansatzes.
The plot shows the calculated ground state energy from the ansatz for progressively larger pruning ratios.
The red dotted line is the noise-free correct energy for the LiH molecule.
The boxplots are the median, first quartile, third quartile, minimum, and maximum of the calculated energy across ten trials, the variation is partially due to noise modeled as Qiskit's FakeManila backend.
The precision of the calculated energy increases with greater pruning.
The accuracy also tends to improve.
        }
        \label{fig:pruned_HWEA}
\end{figure}

\subsection{Motivation for NISQ Circuit Cutting}

A major concern about quantum circuit cutting is the exponential costs in quantum circuit executions and classical post-processing.
Due to these concerns, some types of quantum circuit have been found to be nonscalable in the circuit cutting strategy~\cite{pawar2023integratedqubitreusecircuit}.
This paper makes a case that quantum circuits for NISQ era VQAs have favorable properties in terms of topology, determinism, and sparsity such that they work well with circuit cutting.

VQAs overcome limitations in the accuracy and capacity of QC by combining the strengths of quantum computers and classical optimization.
The quantum circuits serve to evaluate a high-dimensional objective function parameterized by rotation angles.
A classical optimizer drives the QC executions, tuning the rotation angles to minimize the objective function.
VQAs span the full range of envisioned near-term applications of QCs, including quantum chemistry~\cite{Peruzzo2014,QChemQC,RevModPhys.92.015003,gushuISCA21quantumchemistry,Tetris}, optimization~\cite{farhi2014quantum,10.1145/3623278.3624751}, and machine learning classification~\cite{QuantumNAS}.

It has been observed that VQA ansatz circuits have topologies that are suitable for circuit cutting.
In previous studies, the most scalable circuit cutting workloads are usually VQA ansatzes~\cite{CutQC}.
The reason is that ansatzes are usually wide and relatively shallow, meaning they comprise many qubits while having few layers of gates.
Furthermore, qubits are loosely connected with two-qubit interactions that match the loosely connected topology of the prototype QCs~\cite{Kandala2017}.
For example, in two-local hardware efficient ansatz (HWEA) circuits shown in Figure~\ref{fig:HWEA_cut}, the number of cuts needed to partition circuits into a constrained number of physical qubits grows linearly versus both width and depth.
These HWEA circuits can be used in both quantum chemistry and ML classifier workloads~\cite{QuantumNAS}, and will be a representative case study in this paper.

Beyond topology, this paper observes that two properties termed determinism and sparsity make VQA ansatz work well with circuit cutting.
In recent work, researchers found that parameter pruning is an effective strategy for VQAs, as shown in Figure~\ref{fig:pruned_HWEA}.
In a pruning strategy, the rotation angles are tuned to minimize the objective function as is typical for VQA, but as the optimization progresses, the smallest angles are set to zero, in effect removing those rotation gates~\cite{gushuISCA21quantumchemistry,QuantumNAS}.
Pruning is an effective strategy for training classical neural networks~\cite{frankle2018the,li2017pruning,cai2020once} and is also helpful for VQAs in part because it reduces the dimensionality of the optimization problem and noise.

In this work, we argue that pruning confers determinism, which aids in reducing the number of quantum executions needed for a given level of accuracy, and also sparsity, which reduces the classical postprocessing costs.

\subsection{Background on Pauli Operators}
\label{sec:pauli}

Here we review the mathematics specific to the quantum circuit cutting strategy.
In addition to the state vectors and unitary operators that are important in QC~\cite{mike_and_ike,Kaye,mermin_2007,Rieffel}, the circuit cutting strategy is formulated in terms of \emph{expectation values} for \emph{observables} specified as \emph{Pauli strings}.
These types of values are important in circuit cutting as the expectation values are determined one by one instead of all at once, making circuit cutting more tractable than direct quantum circuit simulation.
Furthermore, these types of values are what is gathered from the quantum computer in VQAs.

First, we review quantum \emph{pure} states.
The fundamental unit of computation in a quantum computer is a qubit.
A single qubit state is represented by the vector $\ket{\psi}=\alpha\ket{0}+\beta\ket{1}$.
The \emph{mixed state} is a probabilistic ensemble of pure quantum states.
We use a density matrix to represent the mixed state: $\rho=\sum_{i}p_i\ket{\psi_i}\bra{\psi_i}$, where $\bra{\psi_i}$ is the conjugate transpose of $\ket{\psi_i}$ and $p_i$ is the probability that the mixed state is in the pure state $\ket{\psi_i}$ and $\sum_ip_i=1$. 
The \emph{unitary operation} that applies a unitary gate $U$ to the density matrix description of a quantum state $\rho$ is $\rho^*=U\rho U^\dag$.

The \emph{expectation value} of the state $\rho$ measured on a certain observable $\widehat{O}$ is $\mathrm{tr}(\rho \widehat{O})$, where $\mathrm{tr}$ denotes trace.
The trace is an important operation as it underpins our method of decomposing density matrices.
Here is a clearer explanation of the expectation value ($n$ is \#qubits and $N=2^n$):
{\small\[
\mathrm{tr}(\rho\widehat{O})=\sum_{i=1}^N(\rho\widehat{O})_{ii}\\
=\sum_{i=1}^N\sum_{j=1}^N(\rho_{ij}\widehat{O}_{ji})=\sum_{i=1}^N\sum_{j=1}^N(\rho_{ij}\widehat{O}^{\dag}_{ij})\]}
$\widehat{O}^{\dag}_{ij}$ means the conjugate transpose of $\widehat{O}_{ij}$.

Now we regard the density matrix $\rho$ as a vector of dimension $N^2$.
We can decompose it by $N^2$ \emph{orthogonal bases}.
The $4^n=N^2$ \emph{Pauli strings} are such orthogonal basis vectors.
The Pauli matrices are
{\small\[
I = \begin{bmatrix}
    1 & 0 \\
    0 & 1 \\
\end{bmatrix},
\quad
X = \begin{bmatrix}
    0 & 1 \\
    1 & 0 \\
\end{bmatrix},
\quad
Y = \begin{bmatrix}
    0 & -i \\
    i & 0 \\
\end{bmatrix},
\quad
Z = \begin{bmatrix}
    1 & 0 \\
    0 & -1 \\
\end{bmatrix}.
\]
}
Pauli strings like $XYXZ$ are the Kronecker (tensor) product of these Pauli matrices in the order $X\otimes Y\otimes X\otimes Z$.
The inner product of two different Pauli strings is 0, so the Pauli strings form a complete set of orthogonal bases.

Given a density matrix $\rho$, the way to decompose it is to find the inner product with the orthonormal basis.
Thus, we can write $\rho$ in a new way: $\rho=\sum_{\text{ps}\in\text{Pauli strings}}f(\text{ps})*\text{ps}$, where $\text{ps}$ is one of the Pauli strings like $XY\cdots Z$, and $f(\text{ps})$ is the coefficient corresponding to that Pauli string.
$f(\text{ps})=\mathrm{tr}(\rho\text{ps})/2^n$.
One thing to note, since every density matrix is Hermitian (a complex square matrix that is equal to its own conjugate transpose), and each Pauli string is also Hermitian, so when calculating $f(\text{ps})=\frac{1}{2^n}\sum_{i=1}^N\sum_{j=1}^N(\rho_{ij}\text{ps}_{ij}^{\dag})$, $\rho_{ij}\text{ps}_{ij}^{\dag}$ is the conjugate of $\rho_{ji}\text{ps}_{ji}^{\dag}$ when $i\neq j$.
So, the imaginary part of the coefficient $f(\text{ps})$ is always zero, which means that all coefficients $f$ are real values.

Finally, an example. Consider
$\mathrm{QFT}\ket{11}$.
Its state vector is:
{\small\begin{align*}
\mathrm{QFT}\ket{11}
&=\frac{1}{\sqrt{2}}(\ket{0}-\ket{1})\frac{1}{\sqrt{2}}(\ket{0}-i\ket{1})
=\ket{-}\ket{-i}
\end{align*}}
Its density matrix is therefore: 
{\small\begin{align*}
\ket{-}\bra{-}\otimes\ket{-i}\bra{-i}
&=\frac{1}{4}II-\frac{1}{4}IY-\frac{1}{4}XI+\frac{1}{4}XY
\end{align*}}
The observable Pauli strings are $II, IY, XI, XY$ and their corresponding expectation value coefficients $f$ are $\frac{1}{4}, -\frac{1}{4}, -\frac{1}{4}, \frac{1}{4}$, respectively.
We will be using this example in Section~\ref{sec:sparse_tn}.


\begin{figure*}[t!]
\centering
\begin{subfigure}[b]{0.49\linewidth}
\includegraphics[width=\linewidth]{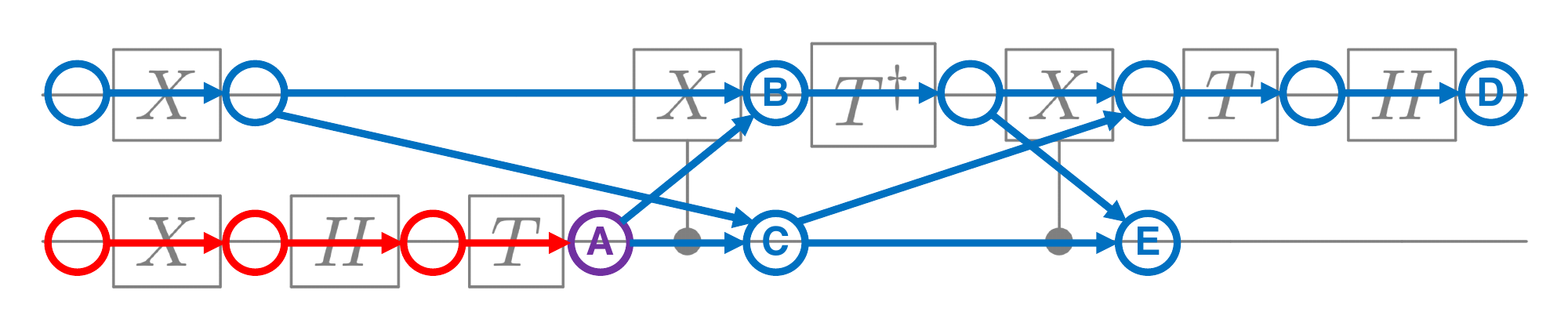}
\caption{
The QFT operation is applied to an $\ket{11}$ initial state set by the $X$ gates.
Two-qubit QFT is non-Clifford, as it requires a controlled-$S$ gate.
The circuit has been decomposed into a Clifford+$T$ gate basis where the only two-qubit gates are $CNOT$ gates.
}
\label{fig:qft2_bn}
\end{subfigure}
\hfill
\begin{subfigure}[b]{0.49\linewidth}
\includegraphics[width=\linewidth]{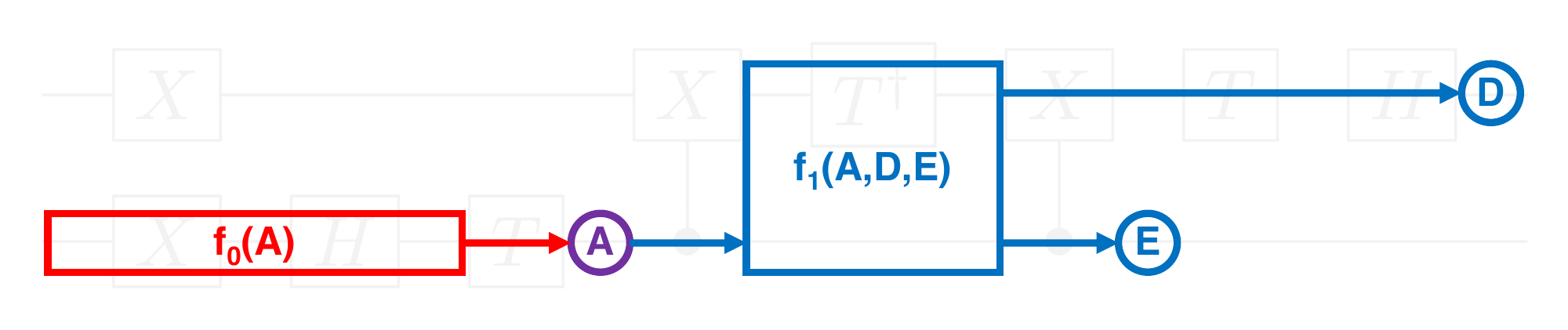}
\caption{
Reconstruction is mathematically equivalent to factor graph exact inference where the red subcircuit forms one factor, the blue subcircuit is another factor, node $A$ is a variable to be eliminated, and nodes $D$ and $E$ are the final query variables.
}
\label{fig:qft2_fg}
\end{subfigure}
\caption{
Directed graphical model representation of cutting, evaluating, and postprocessing a 2-qubit QFT circuit.
}
\label{fig:qft2}
\end{figure*}

\begin{figure*}[t!]
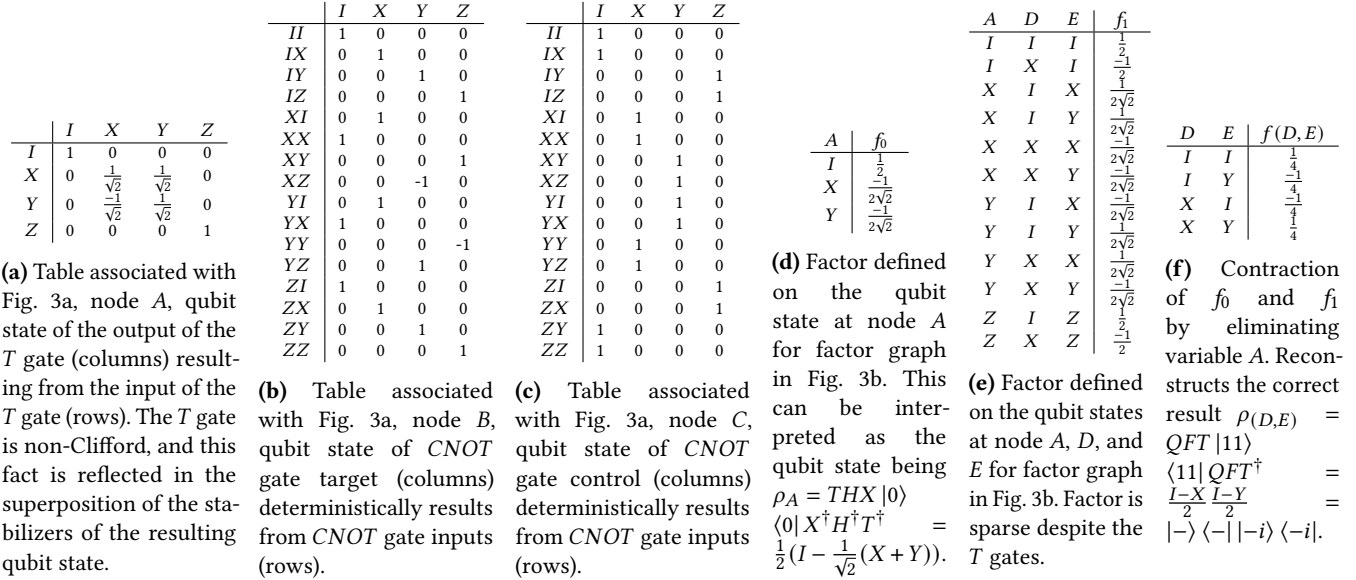

\begin{subfigure}[t]{0.175\linewidth}
\centering
\scriptsize
\begin{tabular}{c|cccc}
& $I$ & $X$ & $Y$ & $Z$ \\\hline
$I$ & 1 & 0 & 0 & 0 \\
$X$ & 0 & $\frac{1}{\sqrt{2}}$ & $\frac{1}{\sqrt{2}}$ & 0 \\
$Y$ & 0 & $\frac{-1}{\sqrt{2}}$ & $\frac{1}{\sqrt{2}}$ & 0 \\
$Z$ & 0 & 0 & 0 & 1 \\
\end{tabular}
\caption{
Table associated with Fig.~\ref{fig:qft2_bn}, node $A$, qubit state of the output of the $T$ gate (columns) resulting from the input of the $T$ gate (rows).
The $T$ gate is non-Clifford, and this fact is reflected in the superposition of the stabilizers of the resulting qubit state.
}
\label{tab:node_A_table}
\end{subfigure}
\hfill
\begin{subfigure}[t]{0.175\linewidth}
\centering
\scriptsize
\begin{tabular}{c|cccc}
& $I$ & $X$ & $Y$ & $Z$ \\\hline
$II$ & 1 & 0 & 0 & 0 \\
$IX$ & 0 & 1 & 0 & 0 \\
$IY$ & 0 & 0 & 1 & 0 \\
$IZ$ & 0 & 0 & 0 & 1 \\
$XI$ & 0 & 1 & 0 & 0 \\
$XX$ & 1 & 0 & 0 & 0 \\
$XY$ & 0 & 0 & 0 & 1 \\
$XZ$ & 0 & 0 & -1 & 0 \\
$YI$ & 0 & 1 & 0 & 0 \\
$YX$ & 1 & 0 & 0 & 0 \\
$YY$ & 0 & 0 & 0 & -1 \\
$YZ$ & 0 & 0 & 1 & 0 \\
$ZI$ & 1 & 0 & 0 & 0 \\
$ZX$ & 0 & 1 & 0 & 0 \\
$ZY$ & 0 & 0 & 1 & 0 \\
$ZZ$ & 0 & 0 & 0 & 1 \\
\end{tabular}
\caption{
Table associated with Fig.~\ref{fig:qft2_bn}, node $B$, qubit state of $CNOT$ gate target (columns) deterministically results from $CNOT$ gate inputs (rows).
}
\label{tab:node_B_table}
\end{subfigure}
\hfill
\begin{subfigure}[t]{0.175\linewidth}
\centering
\scriptsize
\begin{tabular}{c|cccc}
& $I$ & $X$ & $Y$ & $Z$ \\\hline
$II$ & 1 & 0 & 0 & 0 \\
$IX$ & 1 & 0 & 0 & 0 \\
$IY$ & 0 & 0 & 0 & 1 \\
$IZ$ & 0 & 0 & 0 & 1 \\
$XI$ & 0 & 1 & 0 & 0 \\
$XX$ & 0 & 1 & 0 & 0 \\
$XY$ & 0 & 0 & 1 & 0 \\
$XZ$ & 0 & 0 & 1 & 0 \\
$YI$ & 0 & 0 & 1 & 0 \\
$YX$ & 0 & 0 & 1 & 0 \\
$YY$ & 0 & 1 & 0 & 0 \\
$YZ$ & 0 & 1 & 0 & 0 \\
$ZI$ & 0 & 0 & 0 & 1 \\
$ZX$ & 0 & 0 & 0 & 1 \\
$ZY$ & 1 & 0 & 0 & 0 \\
$ZZ$ & 1 & 0 & 0 & 0 \\
\end{tabular}
\caption{Table associated with Fig.~\ref{fig:qft2_bn}, node $C$, qubit state of $CNOT$ gate control (columns) deterministically results from $CNOT$ gate inputs (rows).}
\label{tab:node_C_table}
\end{subfigure}
\hfill
\begin{subfigure}[t]{0.13\linewidth}
\centering
\scriptsize
    \begin{tabular}{c|c}
        $A$ & $f_0$ \\\hline
        $I$ & $\frac{1}{2}$\\
        $X$ & $\frac{-1}{2\sqrt{2}}$\\
        $Y$ & $\frac{-1}{2\sqrt{2}}$\\
    \end{tabular}
    \caption{Factor defined on the qubit state at node $A$ for factor graph in Fig.~\ref{fig:qft2_fg}.
    This can be interpreted as the qubit state being $\rho_A=T H X \ket{0} \newline \bra{0}X^\dag H^\dag T^\dag=\frac{1}{2}(I-\frac{1}{\sqrt{2}}(X+Y))$.
    }
    \label{tab:f_0}
\end{subfigure}
\hfill
\begin{subfigure}[t]{0.13\linewidth}
\centering
\scriptsize
    \begin{tabular}{ccc|c}
        $A$ & $D$ & $E$ & $f_1$ \\\hline
        $I$ & $I$ & $I$ & $\frac{1}{2}$\\
        $I$ & $X$ & $I$ & $\frac{-1}{2}$\\
        $X$ & $I$ & $X$ & $\frac{1}{2\sqrt{2}}$\\
        $X$ & $I$ & $Y$ & $\frac{1}{2\sqrt{2}}$\\
        $X$ & $X$ & $X$ & $\frac{-1}{2\sqrt{2}}$\\
        $X$ & $X$ & $Y$ & $\frac{-1}{2\sqrt{2}}$\\
        $Y$ & $I$ & $X$ & $\frac{-1}{2\sqrt{2}}$\\
        $Y$ & $I$ & $Y$ & $\frac{1}{2\sqrt{2}}$\\
        $Y$ & $X$ & $X$ & $\frac{1}{2\sqrt{2}}$\\
        $Y$ & $X$ & $Y$ & $\frac{-1}{2\sqrt{2}}$\\
        $Z$ & $I$ & $Z$ & $\frac{1}{2}$\\
        $Z$ & $X$ & $Z$ & $\frac{-1}{2}$\\
    \end{tabular}
    \caption{Factor defined on the qubit states at node $A$, $D$, and $E$ for factor graph in Fig.~\ref{fig:qft2_fg}.
    Factor is sparse despite the $T$ gates.}
    \label{tab:f_1}
\end{subfigure}
\hfill
\begin{subfigure}[t]{0.13\linewidth}
\centering
\scriptsize
    \begin{tabular}{cc|c}
        $D$ & $E$ & $f(D,E)$\\\hline
        $I$ & $I$ & $\frac{1}{4}$\\
        $I$ & $Y$ & $\frac{-1}{4}$\\
        $X$ & $I$ & $\frac{-1}{4}$\\
        $X$ & $Y$ & $\frac{1}{4}$\\
    \end{tabular}
    \caption{
    Contraction of $f_0$ and $f_1$ by eliminating variable $A$.
    Reconstructs the correct result $\rho_{(D,E)} = QFT\ket{11}\newline\bra{11}QFT^\dag = \frac{I-X}{2}\frac{I-Y}{2}=\ket{-}\bra{-}\ket{-i}\bra{-i}$.
    }
    \label{tab:f_DE}
\end{subfigure}
\caption{Tables and factors associated with the directed graphical model in Fig.~\ref{fig:qft2_bn} and reconstruction factor graph in Fig.~\ref{fig:qft2_fg}.}
\label{fig:factors}
\end{figure*}

\section{Quantum Circuit Cutting:\\Topology, Determinism, and Sparsity}
\label{sec:sparse_tn}


This section gives a tutorial on the circuit cutting problem.
As a contribution from this article, the circuit cutting problem is presented for the first time in the form of a \emph{factor graph} \emph{exact inference}~\cite{Factor_graphs}, a well-established task in classical artificial intelligence~\cite{aima,Pearl,Koller_Friedman,darwiche_2009}.
This is in contrast to previous work on circuit cutting, which has considered the problem directly using quantum circuits~\cite{CutQC,Clifford-based_Circuit_Cutting, Golden_Circuit_Cutting_Points,pawar2023integratedqubitreusecircuit,Gate_Cuts_and_Wire_Cuts,kan2024scalablecircuitcuttingscheduling} and tensor networks~\cite{Large_Small,tang2022scaleqcscalableframeworkhybrid} as the underlying abstraction.
The advantage of factor graph abstraction is that existing classical AI techniques enable automatic reasoning about determinism and sparsity~\cite{d02,kc_map}, two properties that this work exploits to make circuit cutting more efficient.


\subsection{Execution and Reconstruction Costs due to Topology}
\label{sec:topology}

The quantum circuit topology and how the circuit is cut into subcircuits have a significant impact on the benefit of circuit cutting and its costs.

In this work, circuit partitioning and reconstruction are represented as operations on probabilistic graphical models (PGMs).
Previous work has used PGMs such as tensor networks~\cite{biamonte2017tensornetworksnutshell,Markov_tensor,qTorch,huang2020classicalsimulationquantumsupremacy}, Markov networks~\cite{boixo2018simulationlowdepthquantumcircuits}, and Bayesian networks~\cite{asplos_21} to represent circuits.
An example is given in Figure~\ref{fig:qft2}, where the graph nodes represent the qubit states and the graph edges represent the quantum gates that modify the qubit states.
The circuit cutting task is to partition the graph at nodes into different factors so that the subcircuits can be executed separately on QCs, and the results of the subcircuits are recombined on a classical computer.

A good circuit cutting plan confers benefits.
The number of qubits involved in each subcircuit can be kept within the number of qubits available in the quantum computer.
The number of gates involved can also be kept within the coherence time frame of the devices.
The lower width and depth requirements facilitate the mapping and routing compilation of the circuit to the highest-quality devices.

The penalty of the circuit cutting scheme is the number of subcircuit executions and the cost of reconstruction scales naively as $O(4^k)$, where $k$ is the number of cuts to be made, or more specifically the number of nodes representing the qubit states that partition the circuit~\cite{Treewidth,Markov_tensor}.

All prior work on circuit cutting has focused on these costs and benefits related to the quantum circuit topology.

\begin{figure}[t]
    \centering
    \includegraphics[width=\linewidth]{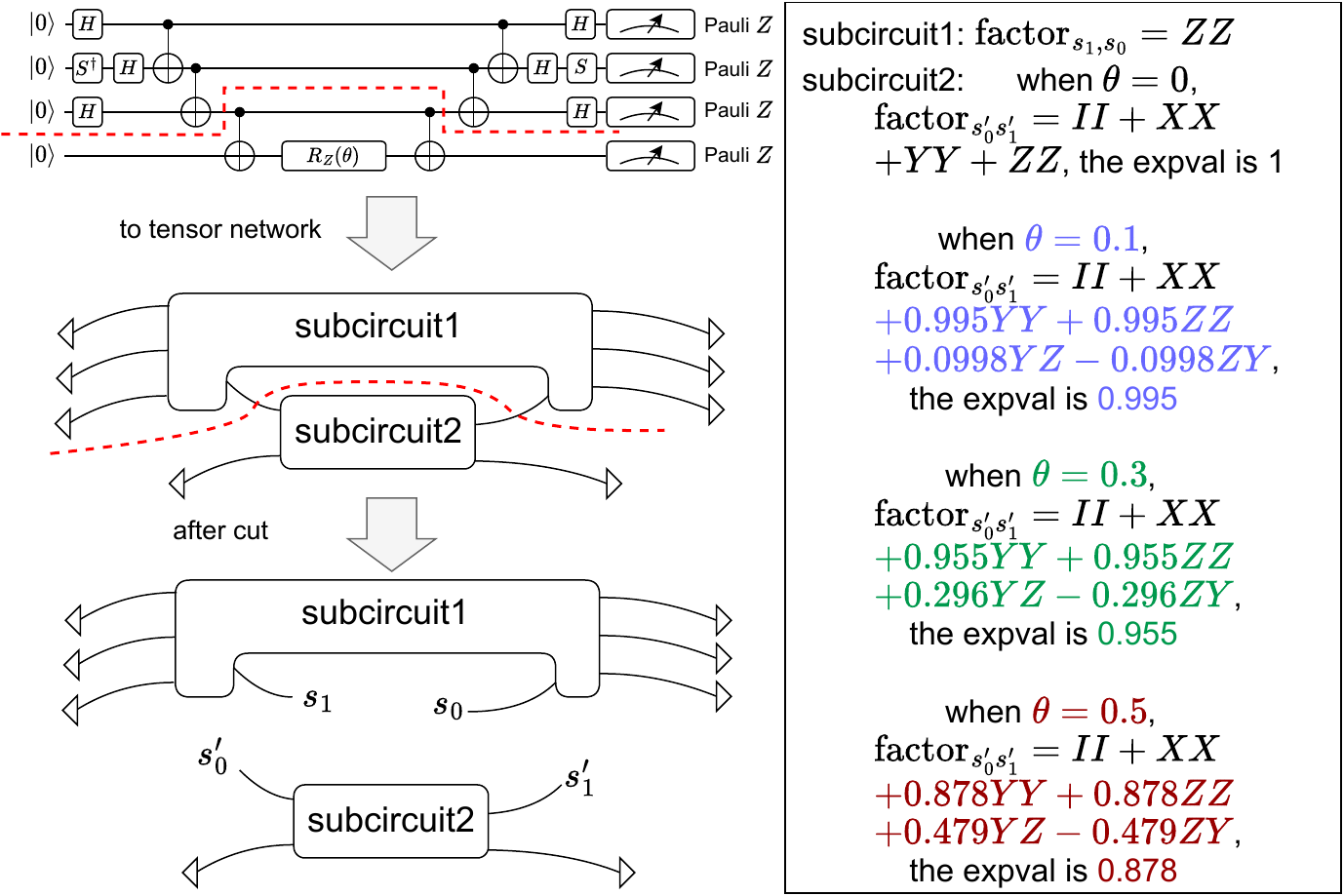}\vspace{-0.0in}
    \caption{Varying non-Clifford rotation gate angles only affects numerical coefficients while the Pauli bases remain deterministically unchanged. The circuit shows a Pauli $XYXZ$ evolution gate on initial state $\ket{0000}$ and measurement of the expectation value on observable $ZZZZ$.}
    \label{fig:subcircuit_function}
\end{figure}

\subsection{Execution Costs due to Determinism}
\label{sec:determinism}

Although previous work in circuit cutting focused on the topology of quantum circuits, the stabilizer and Clifford structure of the circuits are an equally important aspect of the subcircuits execution cost.
The ability to predict, compile, and query this structure is key to this work and leads to efficiencies in the running of QC experiments and to enhancing the accuracy of the results.
The effect is especially relevant to NISQ VQA ansatzes with parameter pruning.

Quantum circuits that contain only stabilized states and Clifford operators are known to be classically easy to simulate in polynomial time and memory requirements~\cite{Heisenberg,simulation_stabilizer,Dominated_by_Clifford,Bravyi2019simulationofquantum}.
In the PGM representation proposed in this paper, the distinction between Clifford and non-Clifford is
equivalent to the concept of \emph{determinism}.
The classic example of determinism in graphical models imagines a scenario in which if a sprinkler is on, the grass is certain to be wet~\cite{aima}.
This knowledge can be captured as a direct entailment without involving any probabilities.
It is a certainty embedded in an otherwise probabilistic model.

In the PGM representation of circuits, Clifford gates are analogous to direct entailment, whereas only non-Clifford gates introduce expectation value coefficients.
For example, the Hadamard gate (which is Clifford) deterministically interchanges states stabilized by the Pauli $X$ and $Z$ operators.
The Clifford two-qubit $CNOT$ gate shown in Figs.~\ref{tab:node_B_table} and~\ref{tab:node_C_table} shows that exactly one stabilizer outcome has a non-zero coefficient for each of the stabilizer inputs.
Therefore, the Clifford Hadamard and $CNOT$ gates represent deterministic transitions between the Pauli bases.
The prototypical non-Clifford $T$ gate shown in Figure~\ref{tab:node_A_table} maps the input Pauli $X$ and $Y$ bases to superpositions of the output Pauli $X$ and $Y$ bases; these mappings introduce non-trivial coefficients.

Classical AI provides a set of \emph{knowledge compilation} techniques that efficiently count, enumerate, and calculate probabilities of events in PGMs~\cite{darwiche_2009,kc_map,KIMMIG201746}.
We apply these techniques to encode determinism, compile causal information, and query this information as needed to drive the circuit cutting strategy.

Knowledge of the Clifford structure can be exploited for more efficient experiment execution.
For example, the subcircuit for $f_1(A,D,E)$ in Figure~\ref{fig:qft2_fg} needs to be executed with various initializations for the qubit state $A$ and measurements on the qubit states $D$ and $E$ in Figure~\ref{tab:f_1}; knowing the valid output states of $A$ from the subcircuit $f_0(A)$ would reveal that experiments that initialize $A$ as Pauli $Z$ are unnecessary.
The reduction in the number of experiments that actually have to be run ripples through the chain of subcircuits.

Knowledge of the Clifford structure can also be exploited to increase the accuracy of reconstructed results.
For example, the possible Pauli strings of the reconstructed result are in Figure~\ref{tab:f_DE} where only four out of 16 bases are non-zero; any strings that do not belong to this set are due to noise, and omitting those strings forms a new kind of error mitigation.

In fact, as shown in Figure~\ref{fig:subcircuit_function}, in VQAs that we claim are especially suited for circuit cutting, only the non-Clifford gates in the ansatz circuit expand the set of Pauli strings for the input and output of the subcircuit.
The variation of the rotation angles $\theta$ on the non-Clifford gates as the variational algorithm executes only amounts to changes in the observable weights of the Pauli strings.
As the angles are increasingly pruned according to Figure~\ref{fig:pruned_HWEA}, the set of nonzero Pauli strings decreases, decreasing the number of experiments needed and strengthening the error mitigation effect.
These ideas are evaluated in Sections~\ref{sec:clever_tomography} and~\ref{sec:error_mitigation}.

\subsection{Reconstruction Costs due to Sparsity}
\label{sec:sparsity}

The fact that the Clifford structure of quantum circuits creates structure in subcircuit inputs and outputs makes the data for the reconstruction postprocessing very sparse.
The sparsity of the data collected from the subcircuit evaluation is clearly seen in Figure~\ref{tab:f_1}.
Of the $4^3=64$ possible Pauli strings for the input and output combinations of the $f_1$ subcircuit, only 12 are consistent with the Clifford structure of the subcircuit and have a non-zero expectation value.

One reason that previous work may have overlooked this sparsity in the data and failed to exploit it may be because the sparsity is only obvious when expressed in the stabilizer Pauli string basis.
Previous work in circuit cutting represented the results of subcircuit executions on the basis of subcircuit input qubit initialization states~\cite{Large_Small}, where the sparsity is not obvious.

Whereas all previous work on quantum circuit cutting has treated this task using conventional matrix and tensor multiplication, a contribution of this work is to process these data as sparse tensor contraction using hashes.
The improved algorithm leads to a reduced number of floating-point operations and a reduced memory footprint needed for post-processing.
These ideas are evaluated in Sections~\ref{sec:flop_count} and~\ref{sec:memory_footprint}.

\section{Efficient Circuit Cutting Execution and Noise Mitigation Exploiting Determinism}
\label{sec:subcircuitevaluation}

Prior work in quantum circuit cutting raised concerns that the number of subcircuit executions is non-scalable.
In this section, we first review how the subcircuit execution cost arises due to the input and output combinations needed to perform quantum state tomography.
That naive baseline approach is what is implemented in standard circuit cutting frameworks such as \texttt{qiskit-addon-cutting}~\cite{qiskit-addon-cutting}.
In NISQ VQA ansatzes with pruned parameters, one can perform significantly better than the baseline.
Once the ansatz topology and cutting plan are set, non-zero observations can be deterministically predicted via extended stabilizer simulation.
Exploiting the determinism leads to either fewer subcircuit executions, tomography results with greater precision, or both.
Finally, we show that our approach matches the error mitigation and efficiency demonstrated in an orthogonal approach termed \emph{shadow tomography}~\cite{Predicting_many_properties, Classical_Shadows}.

\subsection{Background: Execution Inputs and Outputs}
\label{sec:transforming_basis}


For a subcircuit with $n$ input edges and $m$ output edges, it is necessary to try the four initialization states $\{\ket{0}\bra{0},\ket{1}\bra{1},\\\ket{+}\bra{+},
\ket{i}\bra{i}\}$ for each input edge, and for each output edge, it is necessary to try the three observables $\{X,Y,Z\}$ for complete tomography of the quantum state.
Take Figure~\ref{fig:subcircuit_function} for example, where that VQA ansatz circuit partitioning plan has both mid-circuit measurements and mid-circuit initializations.
Subcircuit 1 has an input edge $s_0$ and an output edge $s_1$.
Subcircuit 2 has an input edge $s'_0$ and an output edge $s'_1$.

Therefore, we need to map the subcircuit input Pauli strings $\{I,X,Y,Z\}$ to the actual $\{\ket{0}\bra{0},\ket{1}\bra{1},\ket{+}\bra{+},\ket{i}\bra{i}\}$ states to which qubits can be initialized.
There are $4^{n}3^{m}$ circuit input output settings and, without further knowledge of the ansatz topology, one would have to naively evaluate all the settings on the quantum computer.
Let the function $g$ record the results.
{\small
\begin{gather*}
g(s_0,s_1,\dots,s_{n-1},s_n,\dots,s_{n+m-1})\in [-1, 1],\\
s_i \in \{ \ket{0}\bra{0},\ket{1}\bra{1},\ket{+}\bra{+},\ket{i}\bra{i} \}, 0\leq i<n\\
s_j \in \{I, X, Y, Z\}, n\leq j<n+m.\\
\end{gather*}
}

Notice that $g$ is a linear function, and we have:
{\small
\begin{align*}
I&=\ket{0}\bra{0}+\ket{1}\bra{1}\\
X&=2\ket{+}\bra{+}-\ket{0}\bra{0}-\ket{1}\bra{1}\\
Y&=2\ket{i}\bra{i}-\ket{0}\bra{0}-\ket{1}\bra{1}\\
Z&=\ket{0}\bra{0}-\ket{1}\bra{1}
\end{align*}
}
As a result, we can let $f$ be a new function that transforms $s_0$ to the $\{I, X, Y, Z\}$ basis.
Let $f(s_0,\dots)$ mean $f(s_0, s_1,\dots,s_{n+m-1})$:
{\small
\begin{align*}
f(I, \dots)&=g(\ket{0}\bra{0}, \dots)+g(\ket{1}\bra{1}, \dots)\\
f(X ,\dots)&=2g(\ket{+}\bra{+}, \dots)-g(\ket{0}\bra{0}, \dots)-g(\ket{1}\bra{1}, \dots)\\
f(Y, \dots)&=2g(\ket{i}\bra{i}, \dots)-g(\ket{0}\bra{0}, \dots)-g(\ket{1}\bra{1}, \dots)\\
f(Z, \dots)&=g(\ket{0}\bra{0}, \dots)-g(\ket{1}\bra{1}, \dots)
\end{align*}
}

Now we have successfully transformed the first dimension.
Repeating this for the other $n-1$ dimensions, we will get a function $f$ with every $s_i\in\{I, X, Y, Z\}$. 
$f$ is the factor for the entire subcircuit that we discussed in Section~\ref{sec:pauli} and Figures~\ref{tab:f_0},~\ref{tab:f_1}, and~\ref{tab:f_DE}.
We record the function $f$ as a tensor of dimensions $n+m$.
The extent of each dimension is four.
In each iteration, the time complexity of transforming the basis is linear to the size of the tensor, $O(4^{n+m-1})$.
There are $n$ iterations, so the time complexity of performing the entire basis transformation is $O(n4^{n+m-1})$.
This represents the baseline cost of subcircuit execution in circuit cutting~\cite{Large_Small}.

\begin{figure*}
    \begin{subfigure}[b]{0.32\linewidth}
        \centering
        \includegraphics[width=\linewidth]{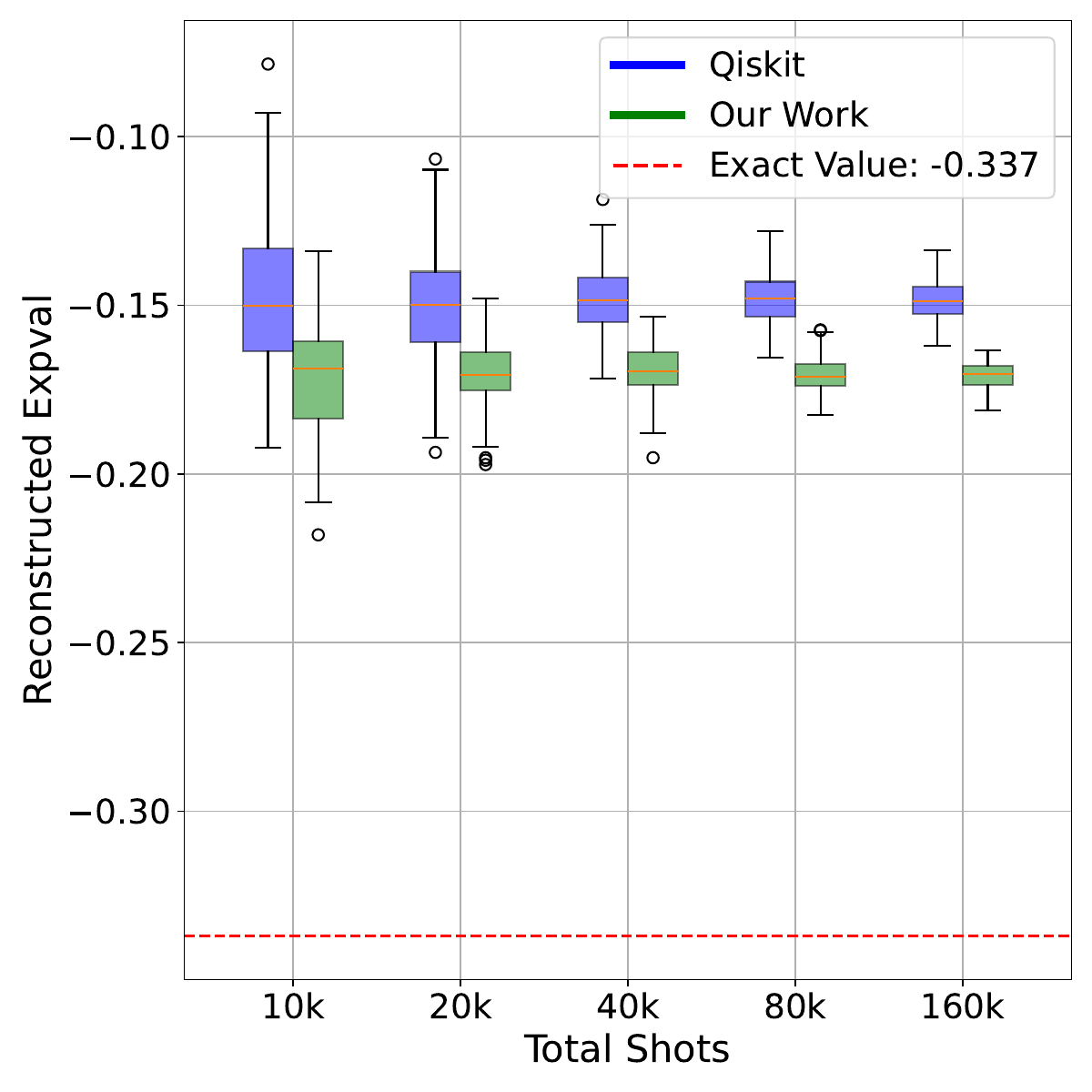}
        \caption{Non-pruned HWEA (8, 1, 5)}
        \label{fig:0_pruned_HWEA(8,1,5)}
    \end{subfigure}
    \hfill
    \begin{subfigure}[b]{0.32\linewidth}
        \centering
        \includegraphics[width=\linewidth]{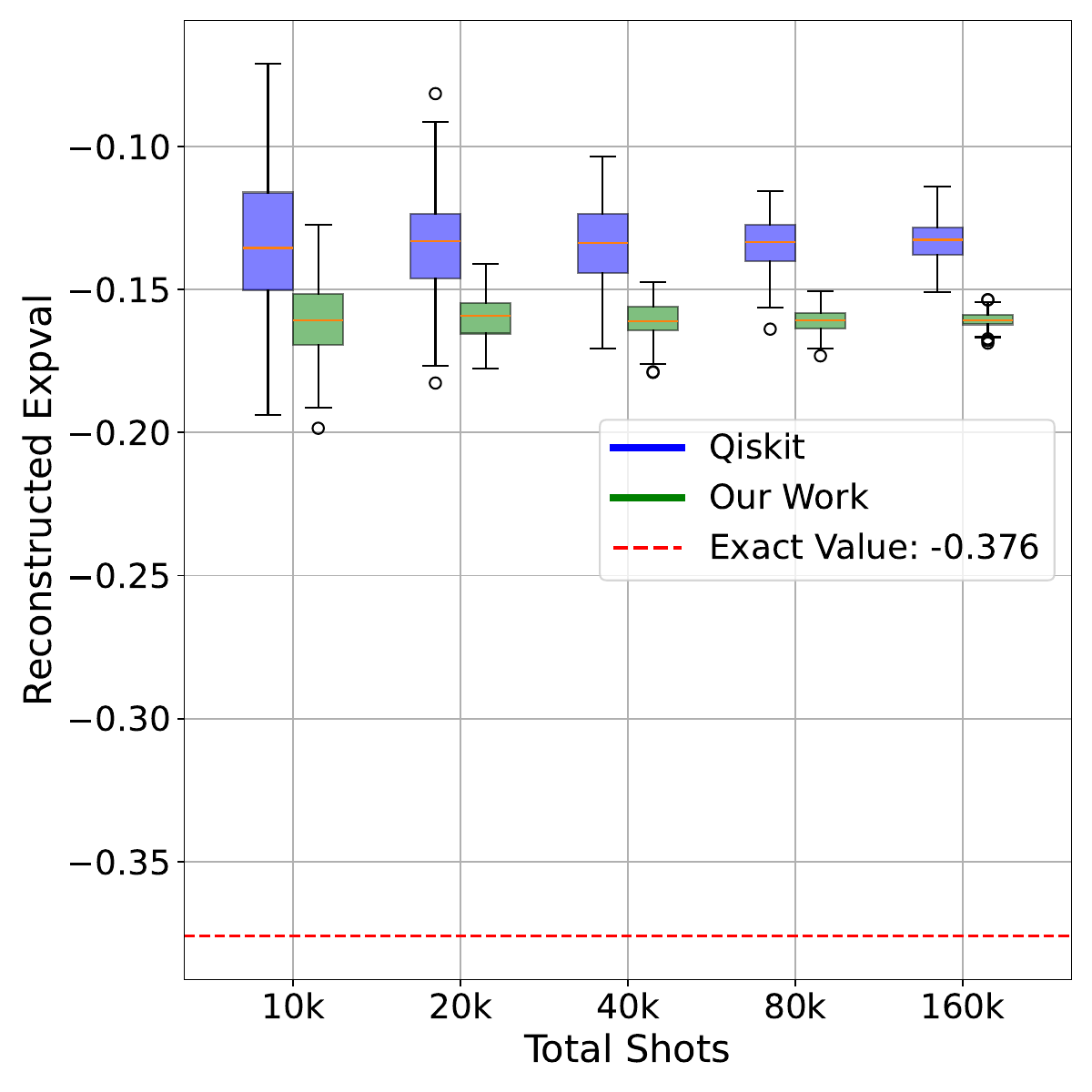}
        \caption{50\%-pruned HWEA (8, 1, 5)}
        \label{fig:0.5_pruned_HWEA(8,1,5)}
    \end{subfigure}
    \hfill
    \begin{subfigure}[b]{0.32\linewidth}
        \centering
        \includegraphics[width=\linewidth]{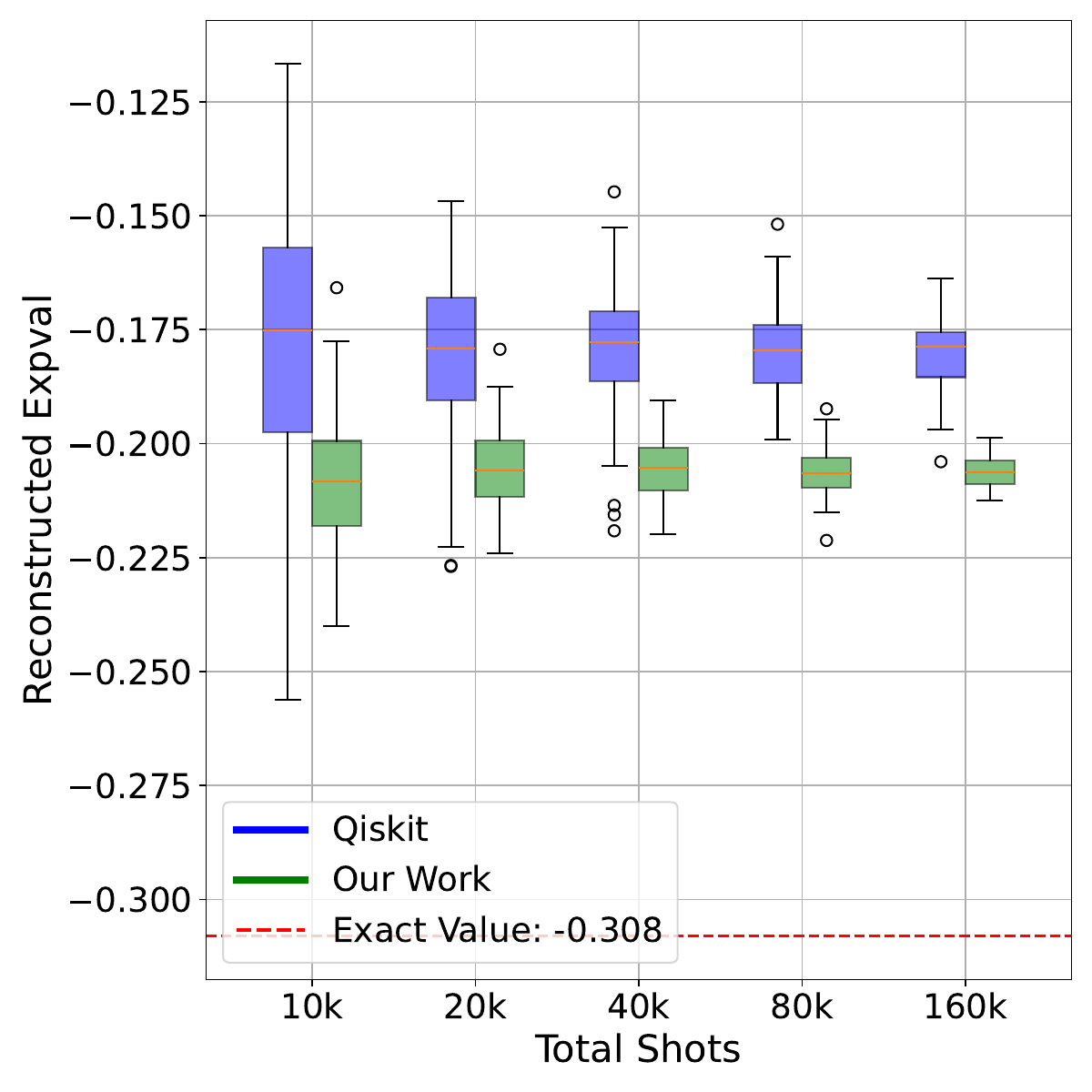}
        \caption{90\%-pruned HWEA (8, 1, 5)}
        \label{fig:0.9_pruned_HWEA(8,1,5)}
    \end{subfigure}
    \begin{subfigure}[b]{0.32\linewidth}
        \centering
        \includegraphics[width=\linewidth]{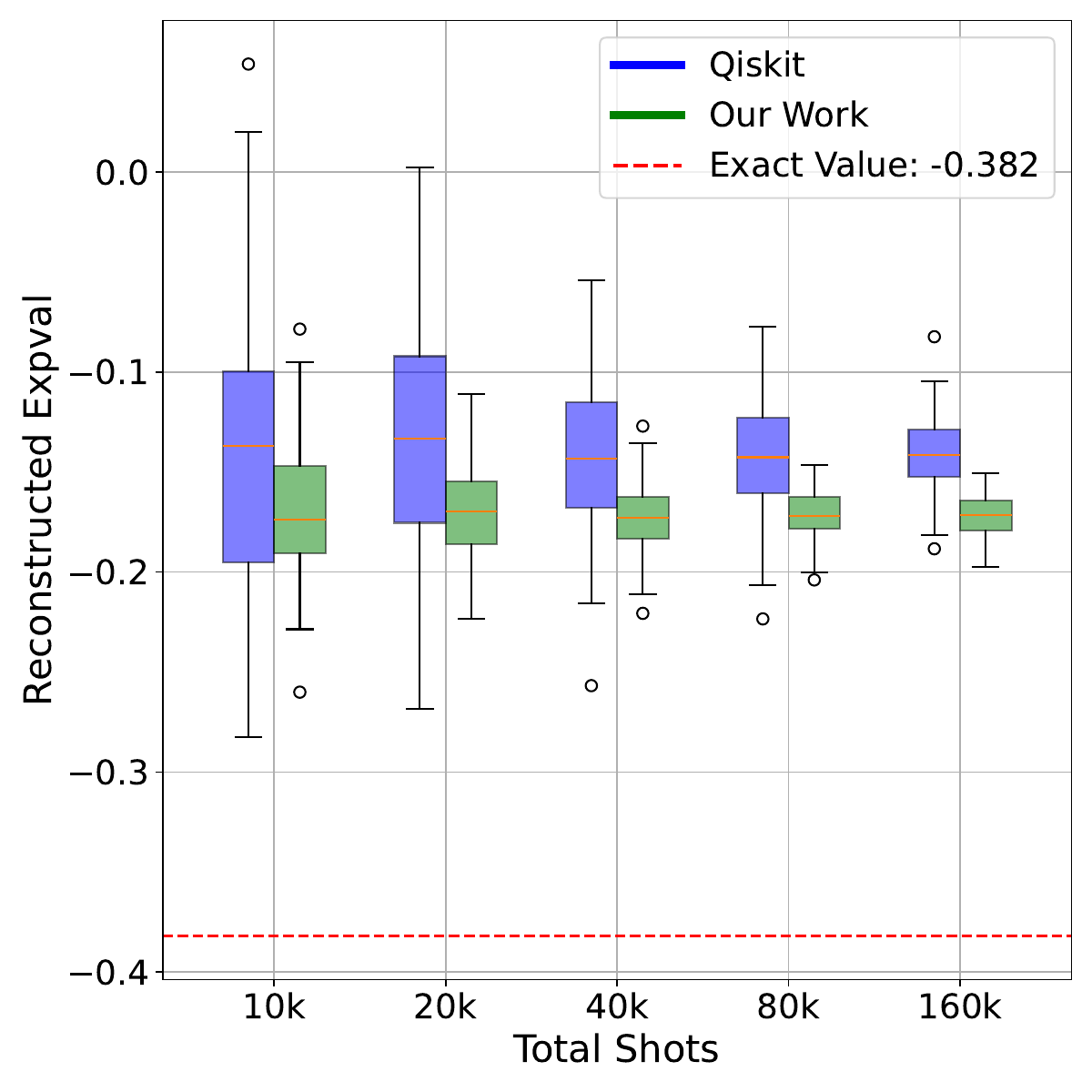}
        \caption{Non-pruned HWEA (8, 2, 5)}
        \label{fig:0_pruned_HWEA(8,2,5)}
    \end{subfigure}
    \hfill
    \begin{subfigure}[b]{0.32\linewidth}
        \centering
        \includegraphics[width=\linewidth]{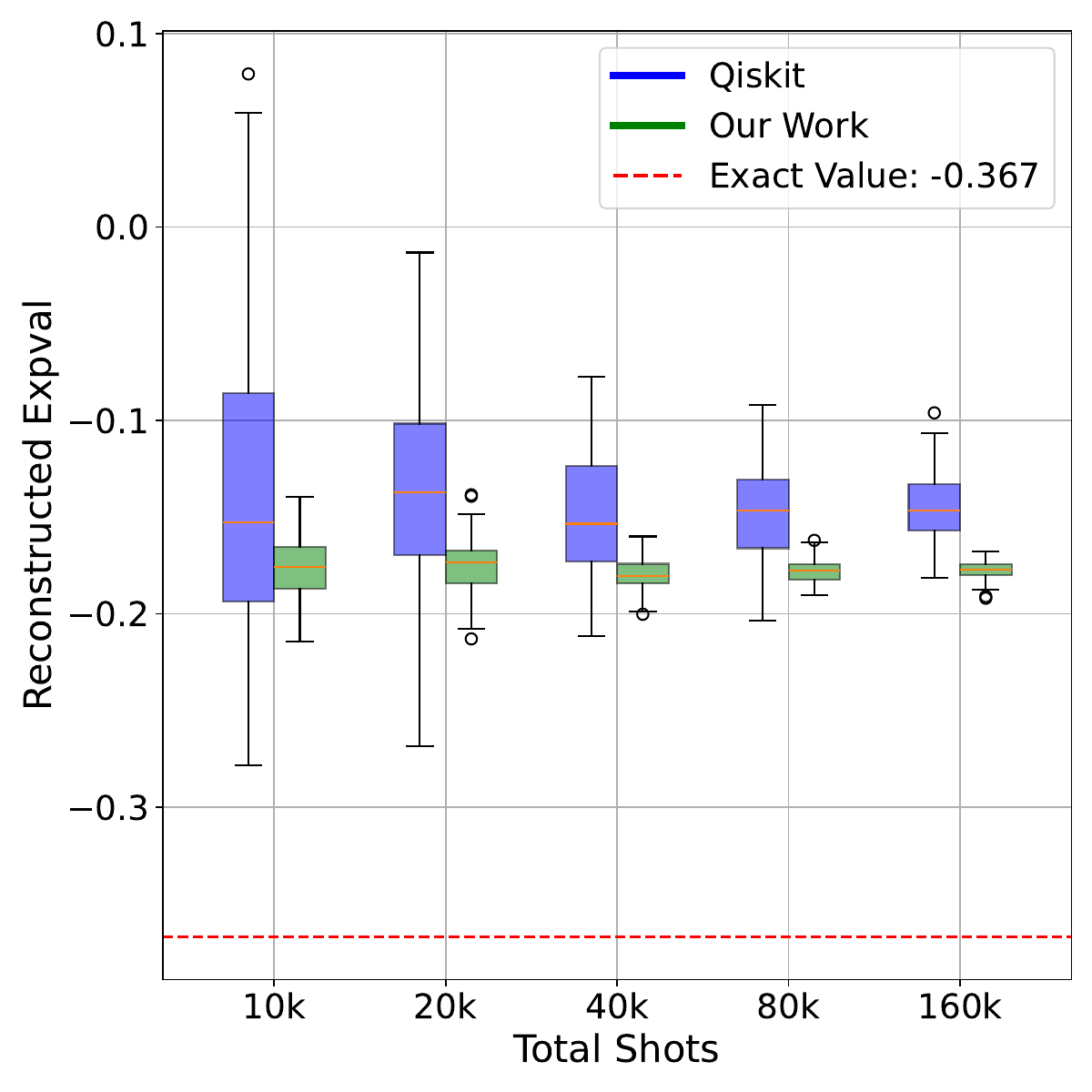}
        \caption{50\%-pruned HWEA (8, 2, 5)}
        \label{fig:0.5_pruned_HWEA(8,2,5)}
    \end{subfigure}
    \hfill
    \begin{subfigure}[b]{0.32\linewidth}
        \centering
        \includegraphics[width=\linewidth]{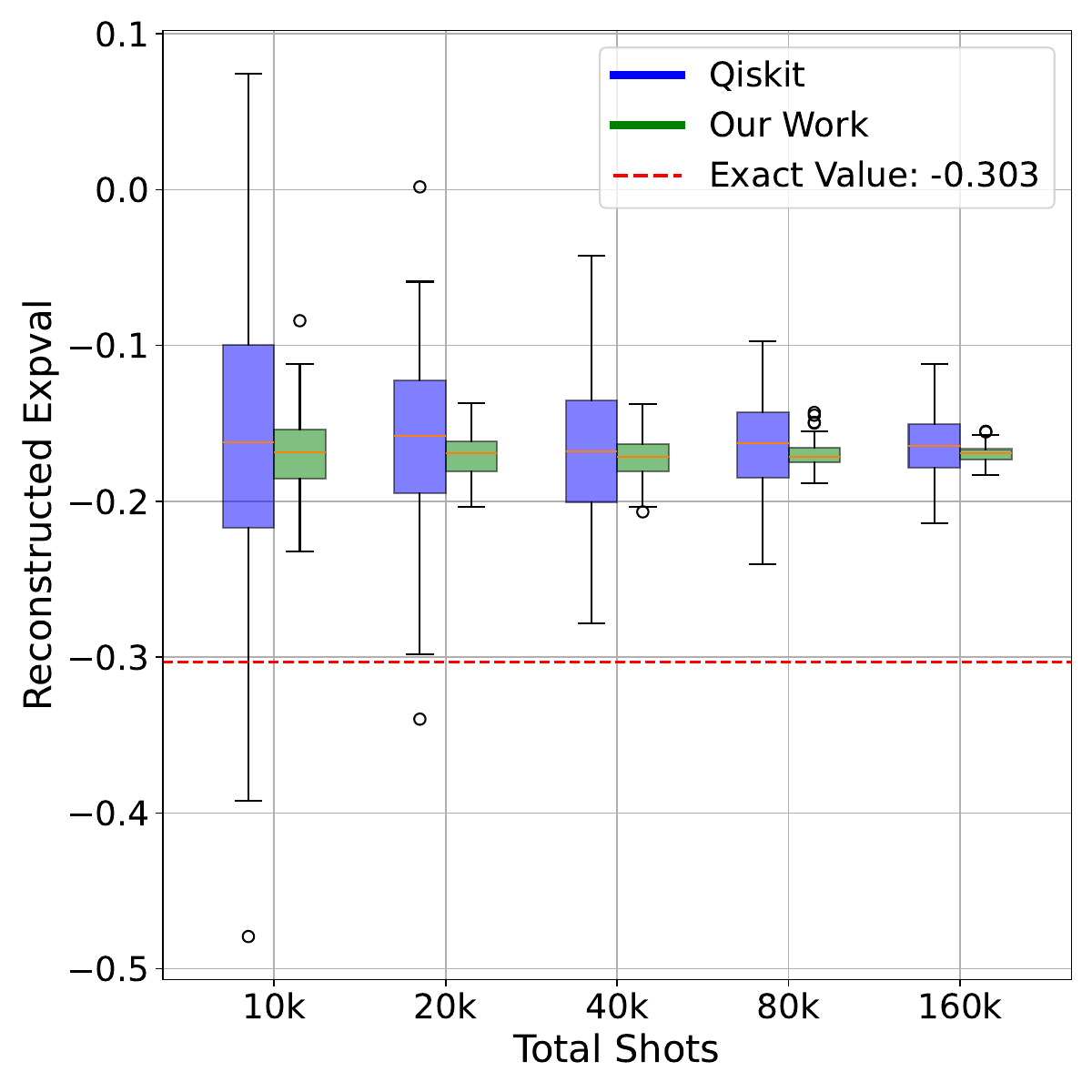}
        \caption{90\%-pruned HWEA (8, 2, 5)}
        \label{fig:0.9_pruned_HWEA(8,2,5)}
    \end{subfigure}
    \begin{subfigure}[b]{0.32\linewidth}
        \centering
        \includegraphics[width=\linewidth]{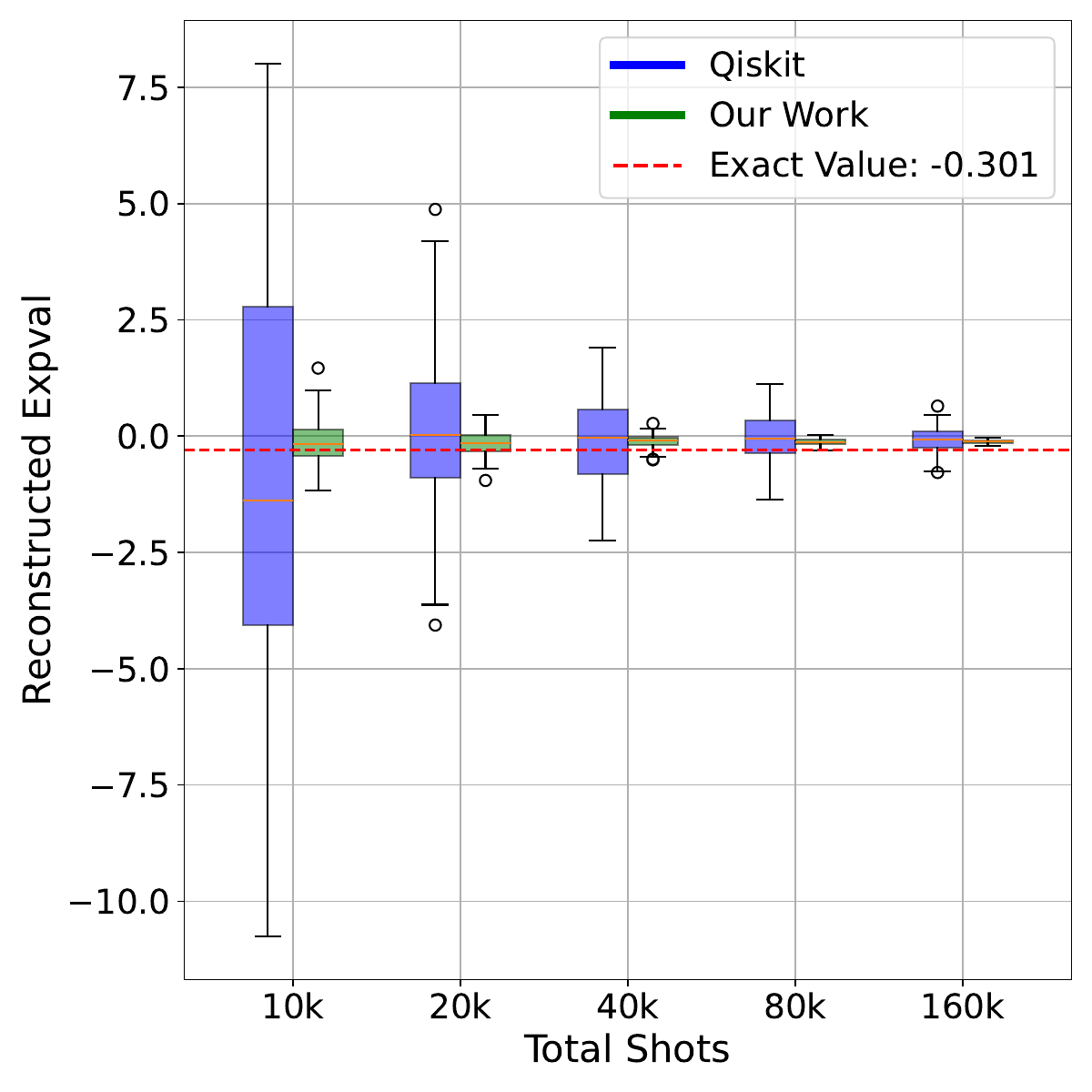}
        \caption{Non-pruned HWEA (8, 4, 5)}
        \label{fig:0_pruned_HWEA(8,4,5)}
    \end{subfigure}
    \hfill
    \begin{subfigure}[b]{0.32\linewidth}
        \centering
        \includegraphics[width=\linewidth]{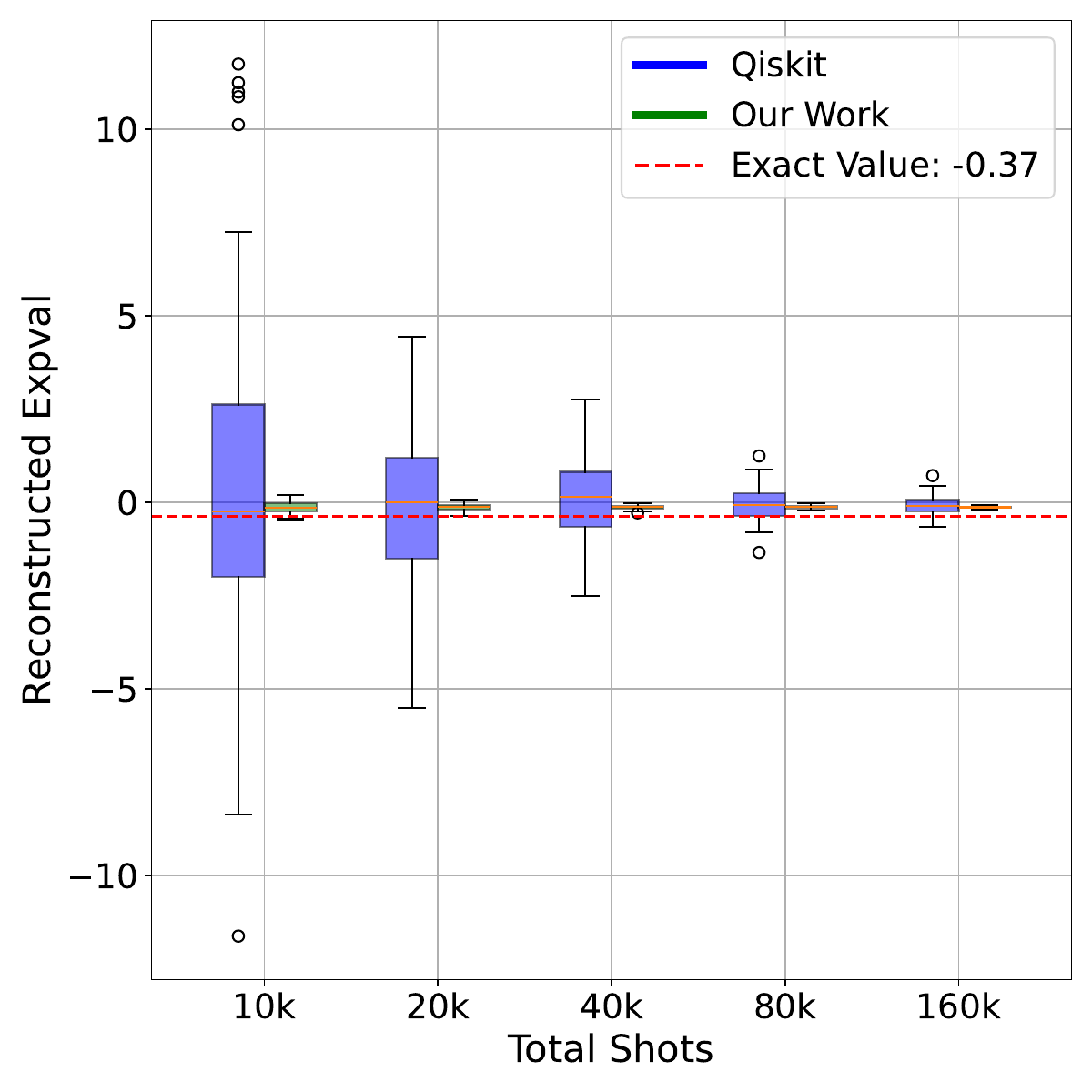}
        \caption{50\%-pruned HWEA (8, 4, 5)}
        \label{fig:0.5_pruned_HWEA(8,4,5)}
    \end{subfigure}
    \hfill
    \begin{subfigure}[b]{0.32\linewidth}
        \centering
        \includegraphics[width=\linewidth]{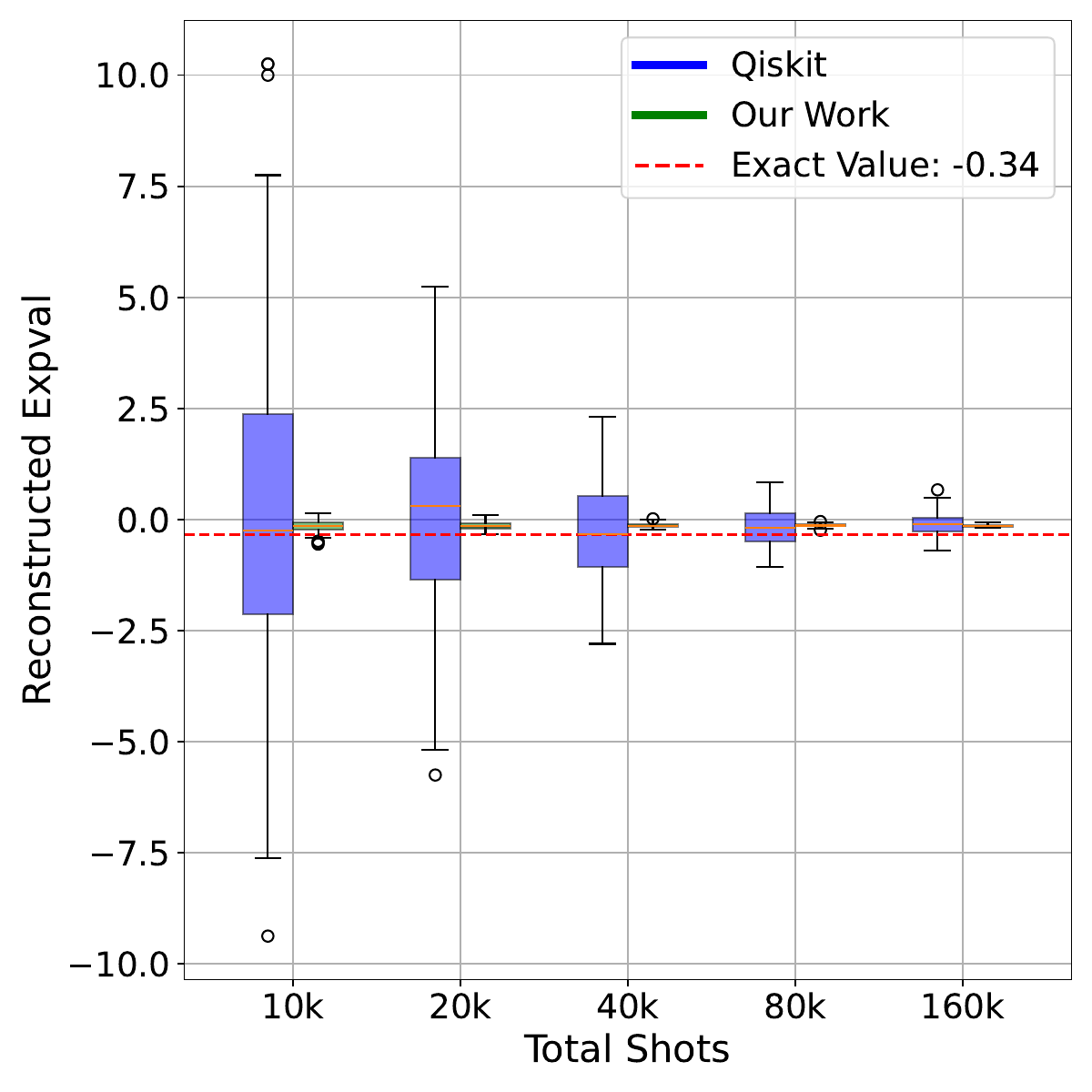}
        \caption{90\%-pruned HWEA (8, 4, 5)}
        \label{fig:0.9_pruned_HWEA(8,4,5)}
    \end{subfigure}

    \caption{
    Efficient subcircuit execution with deterministically known circuit initializations and measurements.
    \emph{In the plots,} the vertical axes are the reconstructed ansatz expectation value.
    The horizontal axes are the total number of shots distributed across input initializations and output measurement combinations.
    The red dotted line is the correct noise-free expectation value for each eight-qubit ansatz circuit.
    The data series are our approach versus the baseline approach for reconstructing the expectation value from the five-qubit subcircuits.
    The boxplots summarize the reconstructed expectation value from 100 trials.
    The variation around the median and the gap from the median to the correct value are due to the noise model from the FakeManila backend in IBM Qiskit.
    \emph{Horizontally across the plots,} the smallest non-Clifford rotation gate angles are progressively pruned as the VQA progresses.
    \emph{Vertically down the plots,} the scheme is tested on two- and four-layer ansatzes.
    }
    \label{fig:depth_pruning_ratio_sensitivity_analysis}
\end{figure*}



\subsection{Efficient Execution with Known Subcircuit Pauli Strings}
\label{sec:clever_tomography}

Here, we exploit knowledge of the correct input and output pairs from the Clifford structure of pruned VQA subcircuits.
Compared to the baseline approach implemented in standard circuit cutting frameworks such as \textit{qiskit-addon-cutting}~\cite{qiskit-addon-cutting}, exploiting this knowledge leads to tomography results with greater precision using fewer subcircuit executions.

\paragraph{Representative VQA workload:}
We study the hardware efficient ansatz consisting of layers of $R_Y$, $R_Z$, and $CZ$ gates shown in Figure~\ref{fig:HWEA_cut} as a standard VQA ansatz that is relevant for both chemistry and quantum machine learning~\cite{Kandala2017}. HWEA(N,d,n) means cutting an N-qubit d-layer HWEA into n-qubit subcircuits.
We keep the qubit count constant at eight qubits and later scale our approach to 200+ qubits in Section~\ref{sec:postprocessing}.
The circuits are cut into subcircuits of width five, such that two cutting points are needed for each ansatz layer.


\paragraph{Baseline versus our approach:}
The \textit{qiskit-addon-cutting} package (we use the latest version 0.9.0 in this paper), previously known as the Circuit Cutting Toolbox, represents the baseline approach that does not know the Clifford structure of the subcircuits.
The baseline approach creates 8 basis elements at each cutting point: $I_0$, $I_1$, $X_0$, $X_1$, $Y_0$, $Y_1$, $Z_0$, $Z_1$.
Therefore, for $k$ cutting points, the total number of experiments is distributed across $8^k$ combinations of input initializations and output measurements.
Once all experiments are executed, the baseline approach reconstructs the quasi-probability decomposition for the subcircuit.

Our work uses the approach described in Section~\ref{sec:determinism} to distribute the total number of experiments over only the subcircuit initializations and measurements that will help the reconstructed state converge to the accurate solution.

As shown in Figure~\ref{fig:depth_pruning_ratio_sensitivity_analysis}, our approach achieves greater precision than the naive approach for a given number of total experiments.
In other words, our approach converges to the same level of precision in fewer total experiments.

\paragraph{Trends across ansatz gate pruning:}
As the pruning ratio grows, the proportion of non-Clifford gates would be smaller, resulting in fewer non-zero Pauli strings in subcircuits.
Our work more efficiently distributes the total number of experiments to fewer correct circuit initialization and measurement combinations, leading to more precise reconstructed values.
The baseline approach cannot take advantage of the fewer combinations that result from gate pruning.

\paragraph{Trends across ansatz layer depth:}
The number of cutting points in the circuit grows linearly with the depth of the ansatz, and the total execution cost of the circuit cutting scheme is understood to be exponential relative to the number of cutting points.
This penalty is obvious for the baseline approach, where for a given budget of the total number of experiments, the limited number of samples for each observable leads to large variances in the reconstructed expectation value.
For four-layer ansatzes shown in Figures~\ref{fig:0_pruned_HWEA(8,4,5)},~\ref{fig:0.5_pruned_HWEA(8,4,5)}, and~\ref{fig:0.9_pruned_HWEA(8,4,5)}, many trials for the baseline approach conclude with an expectation value lower than the correct value due to the large variance, in violation of Ritz's variational principle.
In our approach, deterministic knowledge of the input and output pairs samples only the correct combinations.

\begin{figure}[t]
    \begin{subfigure}[t]{0.48\linewidth}
    \centering
        \includegraphics[width=\linewidth]{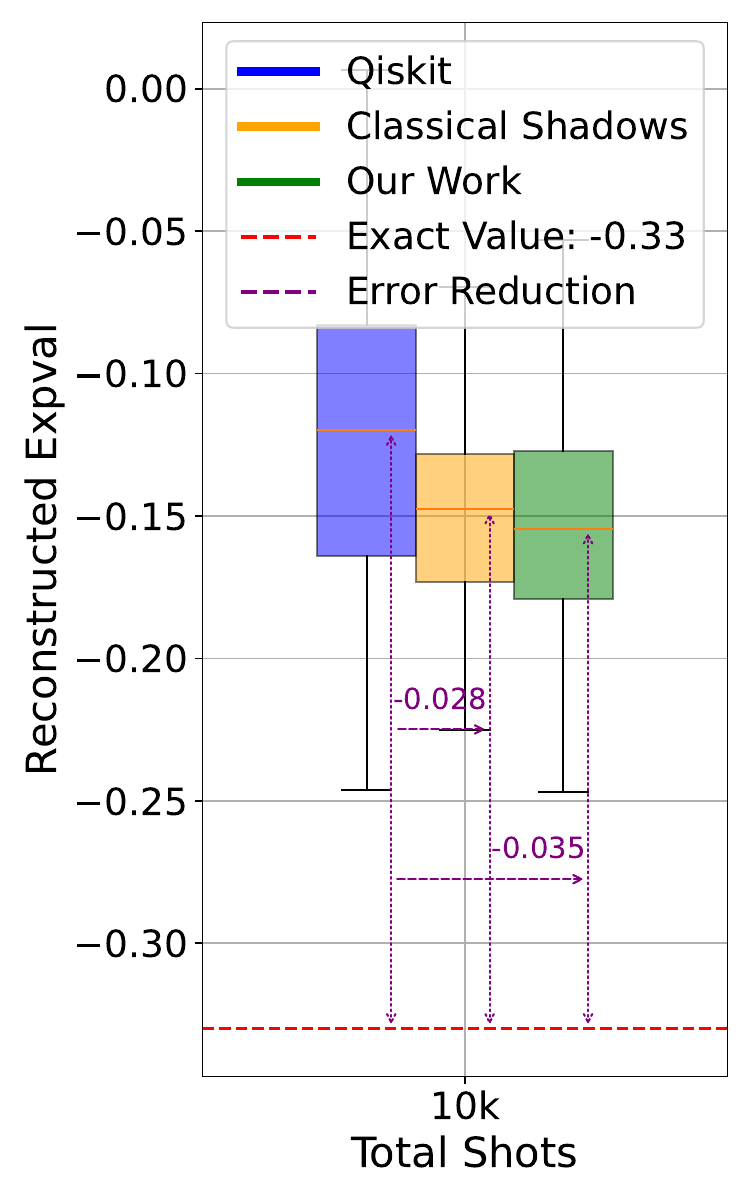}
        \caption{
Non-pruned HWEA (8,2,5).
Each subcircuit involves one upstream cut and one downstream cut.
Since the tensor lacks sparsity, we must conduct all 12 experiments per subcircuit, using 4 bases for the downstream cut and 3 bases for the upstream cut.
Our approach matches the error mitigation of classical shadows in terms of shot efficiency when there is no pruning.
}
        \label{fig:non-pruned-HWEA(8,1,5)-with-classical-shadows}
    \end{subfigure}\hfill
    \begin{subfigure}[t]{0.48\linewidth}
        \centering
        \includegraphics[width=\linewidth]{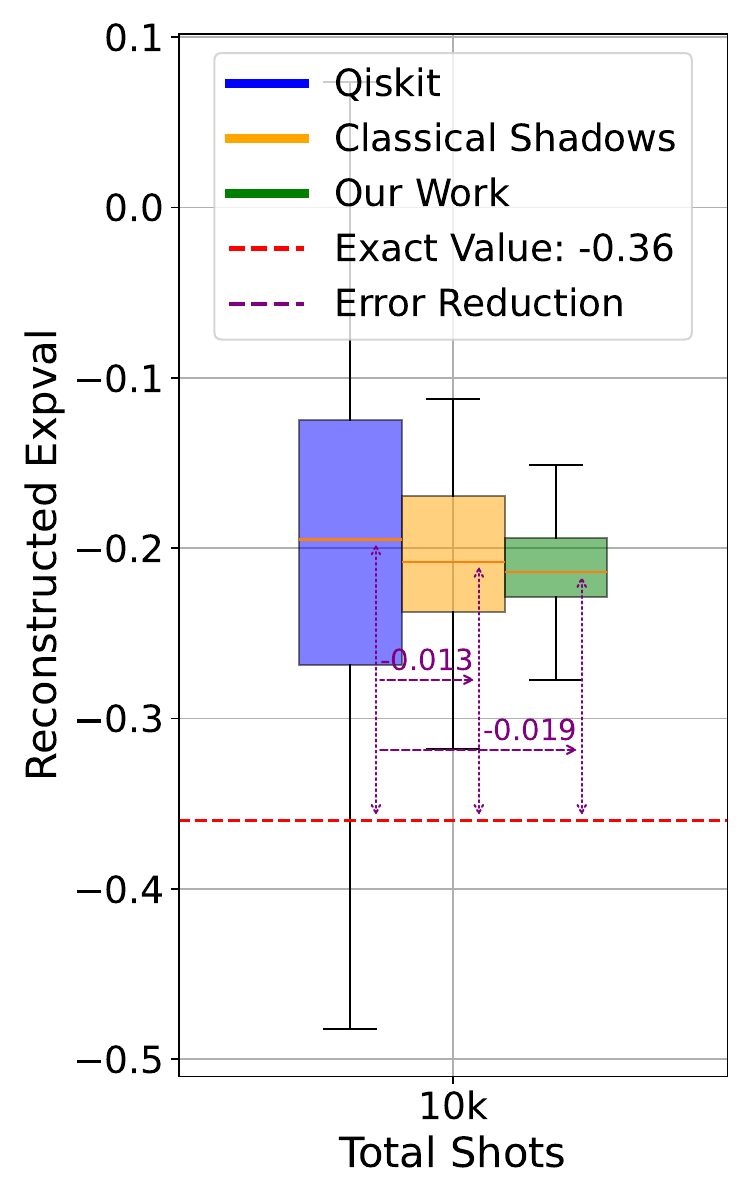}
        \caption{
90\%-pruned HWEA (8,2,5).
Our approach uses determinism to avoid examining every basis.
This reduces the combinations we try to just 12, as opposed to 24, thereby doubling the number of shots per combination compared to the non-pruned scenario.
The denser sampling in our method converges to greater precision in comparison to classical shadows.
}
        \label{fig:0.9-pruned-HWEA(8,1,5)-with-classical-shadows}
    \end{subfigure}
    \caption{
Comparison of the error mitigation capabilities of our approach, classical shadows, and \textit{qiskit-addon-cutting}. 
We limited our tests to a total of $10k$ shots, and the boxplot definitions are consistent with those in Figure~\ref{fig:pruned_HWEA}.
Purple arrows indicate the gap between the median values and the noise-free correct value, representing the absolute error. Both our method and classical shadows achieve an equivalent level of error reduction, yielding more accurate results than \textit{qiskit-addon-cutting}.
}
    \label{fig:classical_shadow}
\end{figure}

\subsection{Error Mitigation with Known Subcircuit Pauli Strings}
\label{sec:error_mitigation}

As seen in Figure~\ref{fig:depth_pruning_ratio_sensitivity_analysis}, the medians of our reconstructed expectation values are closer to the true value when compared against \textit{qiskit-addon-cutting}.
This means that our approach has greater accuracy, in addition to greater precision, than the baseline approach.
In this section, we show that our accuracy improvement matches an orthogonal quantum state tomography method termed classical shadow tomography~\cite{Classical_Shadows,Predicting_many_properties}.
We discuss the source of the error mitigation.

\paragraph{VQA workload:}
We focus this error mitigation study on an eight-qubit two-layer HWEA circuit.
The restriction of the experiment to this circuit is due to the difficulty in simulating the classical shadow tomography procedure.
Although both our approach and \textit{qiskit-addon-cutting} can batch the simulation shots into the subcircuit input and output combinations, the classical shadow tomography approach selects a new input initialization and output measurement basis after every single measurement.


\paragraph{Baseline versus our approach:}


Classical shadow tomography approximates the quantum state density matrix using a few shots by randomly selecting a Pauli basis for each shot, unlike standard tomography that measures on every Pauli basis.
This process is similar to applying an invertible depolarizing channel.
The average density matrix estimated from this method aligns with the true density matrix, allowing for accurate estimation with minimal shots.
However, with a large number of shots, it offers no advantage over standard tomography.
While the classical shadows approach is state-of-the-art, 
it still measures observables with zero expectation value, and those shots contribute nothing to state reconstruction.
Our approach elides those empty shots.

\paragraph{Trends across ansatz gate pruning:}
In Figure~\ref{fig:classical_shadow}, our work and the classical shadow approach are at the same level of mitigating errors compared \textit{qiskit-addon-cutting}.
When there is no pruning, our work matches the classical shadow tomography in terms of precision.
When there is 90\% pruning, our work beats the classical shadow approach in terms of precision. 

\section{Efficient Circuit Cutting Reconstruction Postprocessing Exploiting Sparsity}
\label{sec:postprocessing}

Prior work in quantum circuit cutting also raised concerns that the cost of reconstructing expectation values across the subcircuits using a classical computer is non-scalable.
In this section, we review how the classical post-processing cost arises due to the need to perform a tensor network contraction.
The difficulty of the contraction task is strongly impacted by the choice of the contraction order.
Prior work focused on using state-of-the-art heuristics for finding a contraction order via libraries such as Cotengra.
The contraction task is then performed using GPUs using libraries such as cuQuantum and cuTENSOR.
In our work, we once again show that in NISQ VQA ansatzes with pruned parameters, one can exploit sparsity in the tensors to beat the baseline.
Exploiting sparsity leads to reductions in the operations needed for reconstruction, reduces the memory footprint, and expands the size of NISQ VQA circuits suitable for circuit cutting to sizes previously thought to be intractable.

\begin{figure}[t]
    \centering
    \includegraphics[width=1.0\linewidth]{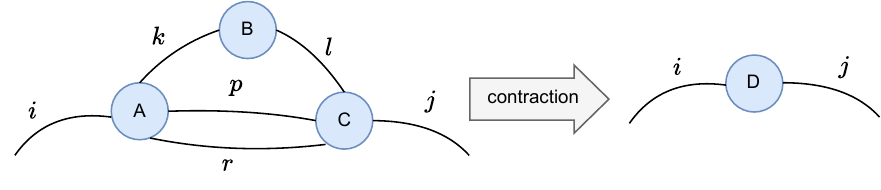}\vspace{-0.0in}
    \caption{Example of tensor contraction.}
    \label{fig:tensor_contraction}
\end{figure}

\subsection{Background: Reconstruction Tensor Contraction}
\label{sec:tensor_contraction}

Here, we discuss why the tensor contraction task arises.
In the discussion about quantum circuit cutting so far, the quantum computer has returned the expectation values for the Pauli string observables for each subcircuit.
These expectation values correspond to the weights of the factors in the graphical model presented in Figures~\ref{fig:qft2_fg},~\ref{tab:f_0}, and~\ref{tab:f_1}.
The classical host computer now has to eliminate the variables representing the qubit states where the original quantum circuit was partitioned, to reconstruct the expectation values for only the output qubit states represented in Figure~\ref{tab:f_DE}.
In prior work on circuit cutting, expectation values are stored as tensors and the topology of how subcircuits are connected is represented as a tensor network.
Eliminating internal variables from the tensor network is termed tensor network contraction.

Tensor contraction is high-dimensional matrix multiplication.
Take Figure~\ref{fig:tensor_contraction} as an example.
We have three tensors $A$, $B$, and $C$.
After contraction, we get a tensor $D$ that has only two edges $i$ and $j$.
{\small\[D_{ij}=\sum_{k, l, p, r}A_{ikpr}B_{kl}C_{prlj}\]}
Tensor sizes grow exponentially with the number of open edges.
In circuit cutting, the extent of each edge is four representing the four Pauli bases, such that the size of the tensor with $k$ edges is $4^k$.
This represents a naive cost of the tensor contraction task.

Tensor network contraction can be done via the \textit{einsum} operation in NumPy, though the state-of-the-art is to use GPU libraries such as cuTENSOR, or more specifically the cuQuantum library part of Nvidia CUDA-Q~\cite{cuQuantum}.

Notice in the example above that the order in which $A$, $B$, and $C$ are eliminated has no effect on the correctness of the outcome $D$, but it does have a significant effect on the computational cost.
For the much trivial case of matrix chain multiplication, where all tensors are rank-two matrices such that the tensor network is just a linear chain, the optimal chain matrix multiplication task has a log-linear-time algorithm for finding the multiplication order.
The optimal contraction ordering for general tensor networks is in contrast NP-Hard.
Here, we use the state-of-the-art Cotengra library to provide heuristic orderings~\cite{Gray2021hyperoptimized}.

\begin{figure*}
    \begin{subfigure}{0.33\linewidth}
        \centering
        \includegraphics[width=\linewidth]{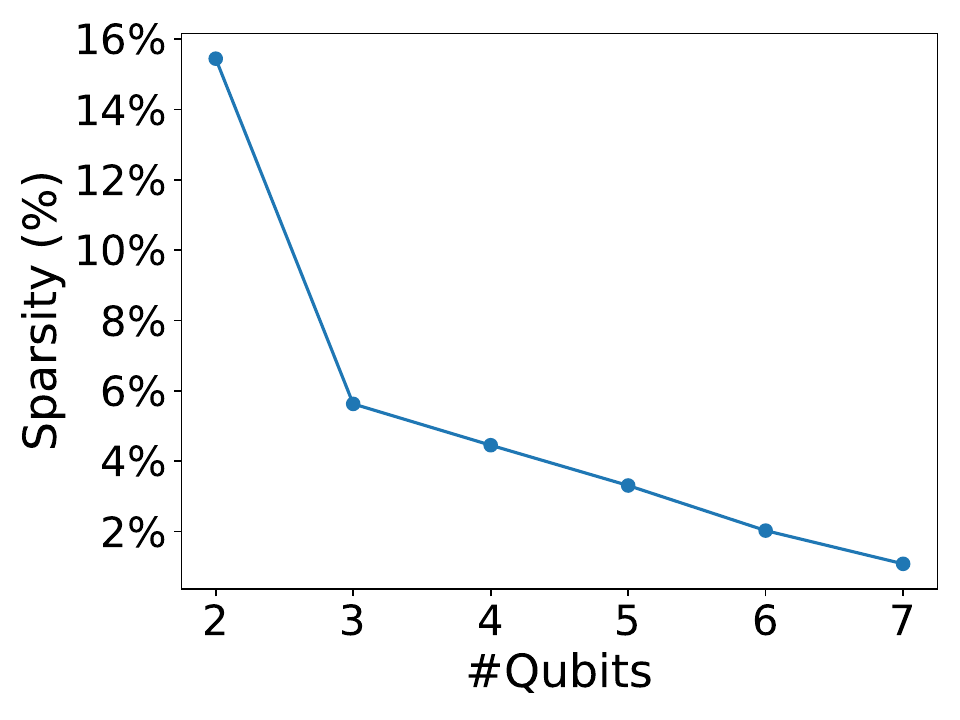}
        \caption{QFT sparsity}
        \label{fig:qft_sparsity}
    \end{subfigure}
    \hfill
    \begin{subfigure}{0.33\linewidth}
        \centering
        \includegraphics[width=\linewidth]{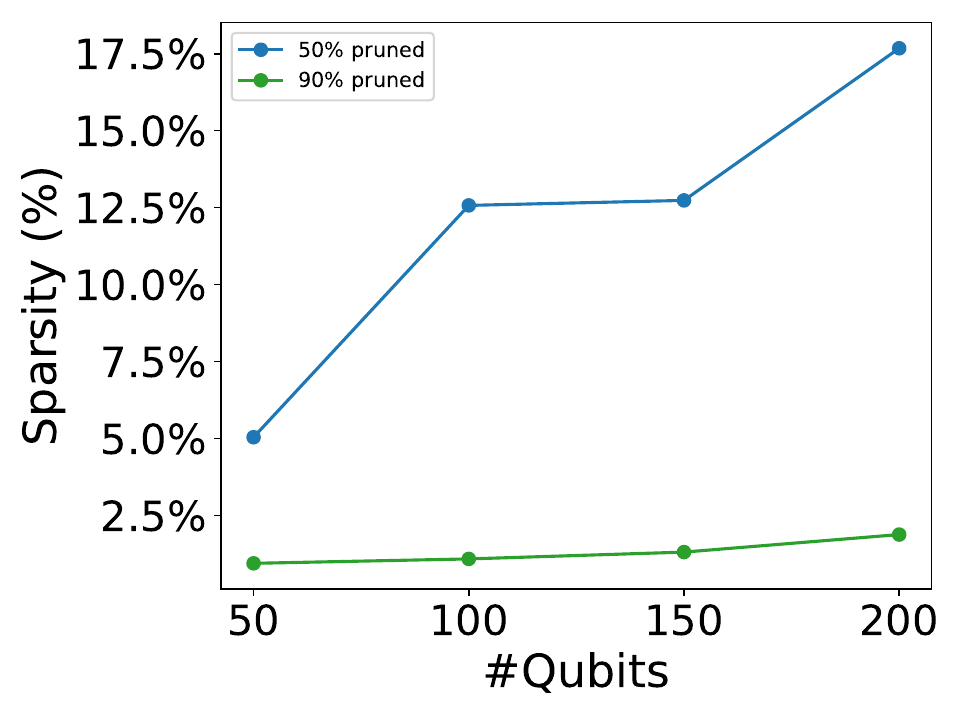}
        \caption{HWEA(\#Qubits,3,20) sparsity}
        \label{fig:HWEA(N,3,20)_sparsity}
    \end{subfigure}
    \hfill
    \begin{subfigure}{0.33\linewidth}
        \centering
        \includegraphics[width=\linewidth]{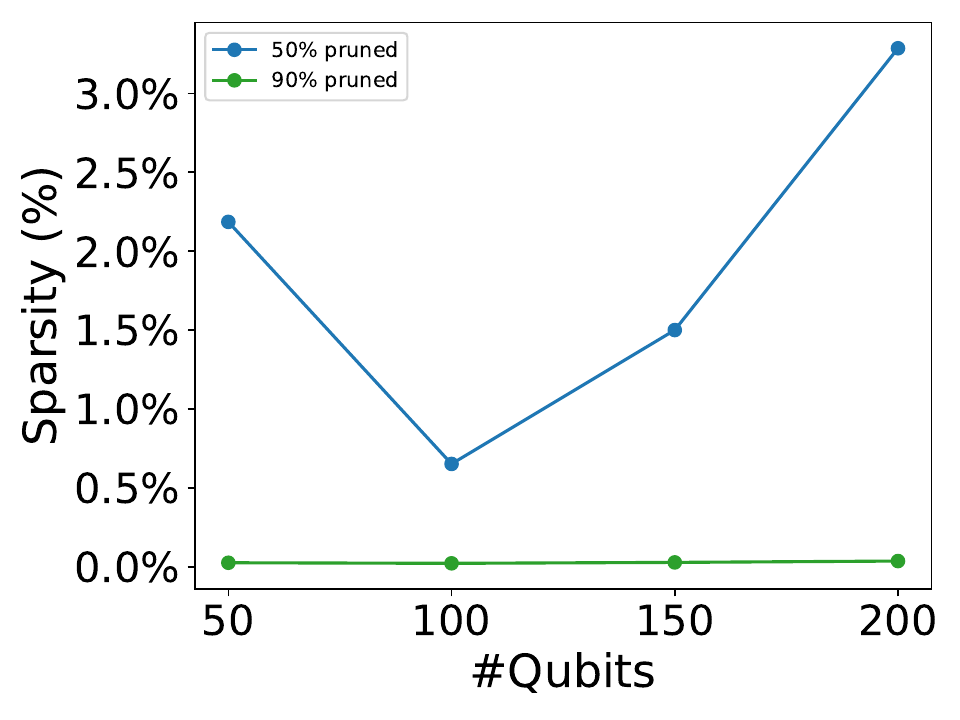}
        \caption{HWEA(\#Qubits,6,20) sparsity}
        \label{fig:HWEA(N,6,20)_sparsity}
    \end{subfigure}
    \begin{subfigure}{0.33\linewidth}
        \centering
        \includegraphics[width=\linewidth]{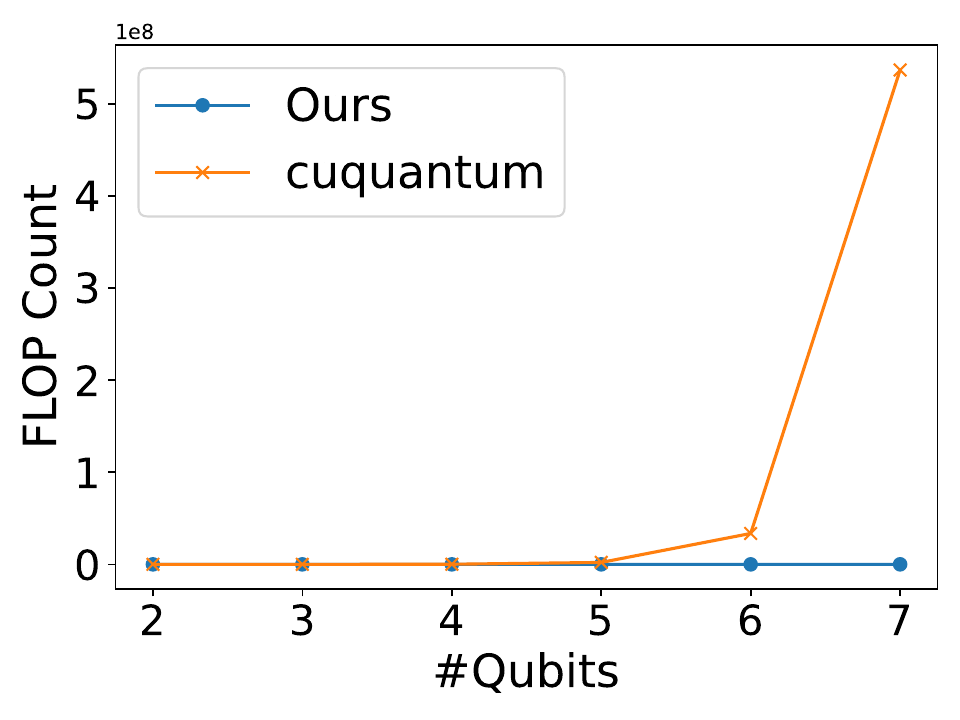}
        \caption{QFT FLOPs}
        \label{fig:qft_flops}
    \end{subfigure}
    \hfill
    \begin{subfigure}{0.33\linewidth}
        \centering
        \includegraphics[width=\linewidth]{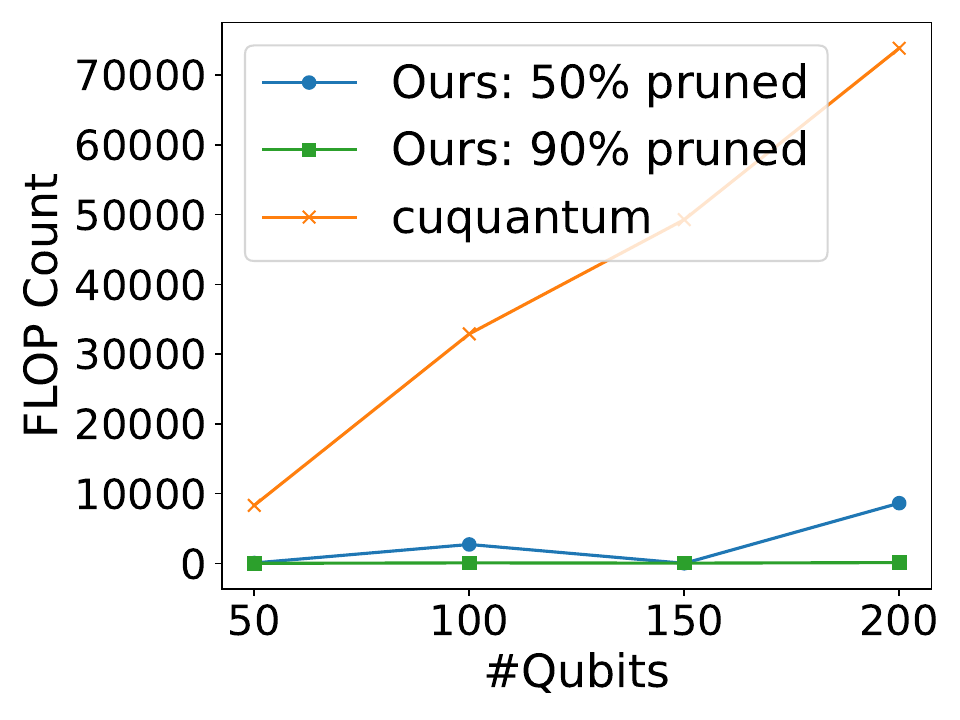}
        \caption{HWEA(\#Qubits,3,20) FLOPs}
        \label{fig:HWEA(N,3,20)_flops}
    \end{subfigure}
    \hfill
    \begin{subfigure}{0.33\linewidth}
        \centering
        \includegraphics[width=\linewidth]{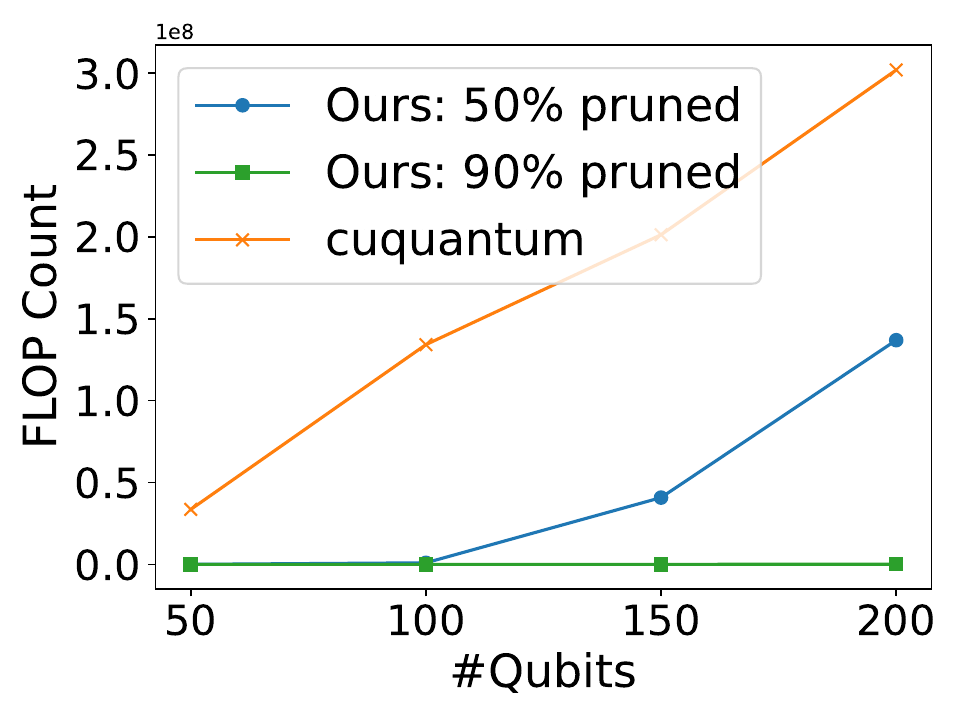}
        \caption{HWEA(\#Qubits,6,20) FLOPs}
        \label{fig:HWEA(N,6,20)_flops}
    \end{subfigure}
    \begin{subfigure}{0.33\linewidth}
        \centering
        \includegraphics[width=\linewidth]{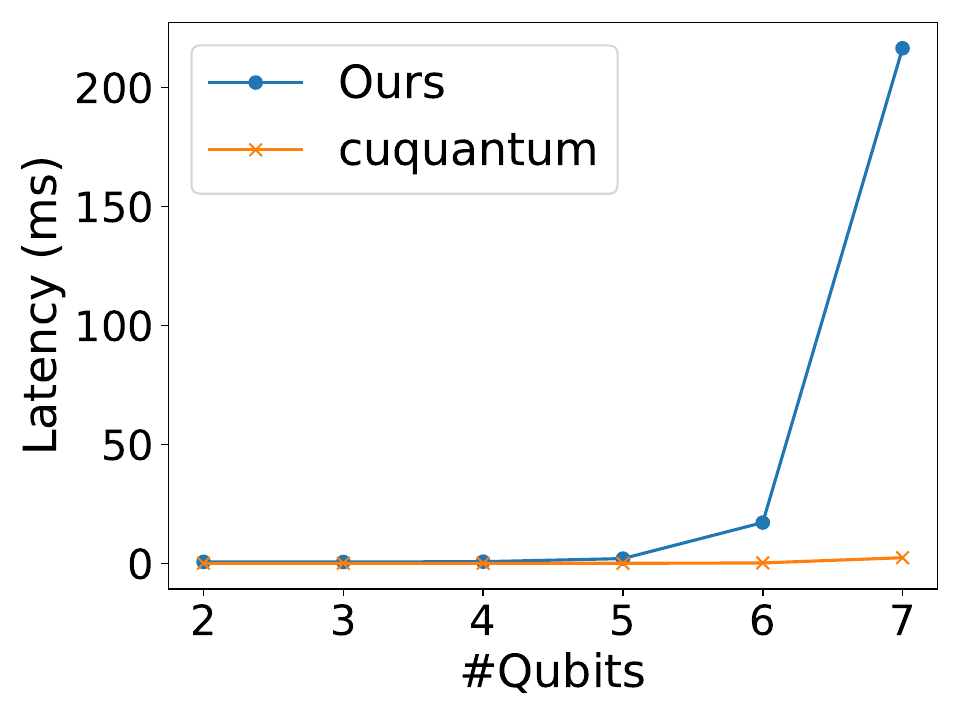}
        \caption{QFT latency}
        \label{fig:qft_latency}
    \end{subfigure}
    \hfill
    \begin{subfigure}{0.33\linewidth}
        \centering
        \includegraphics[width=\linewidth]{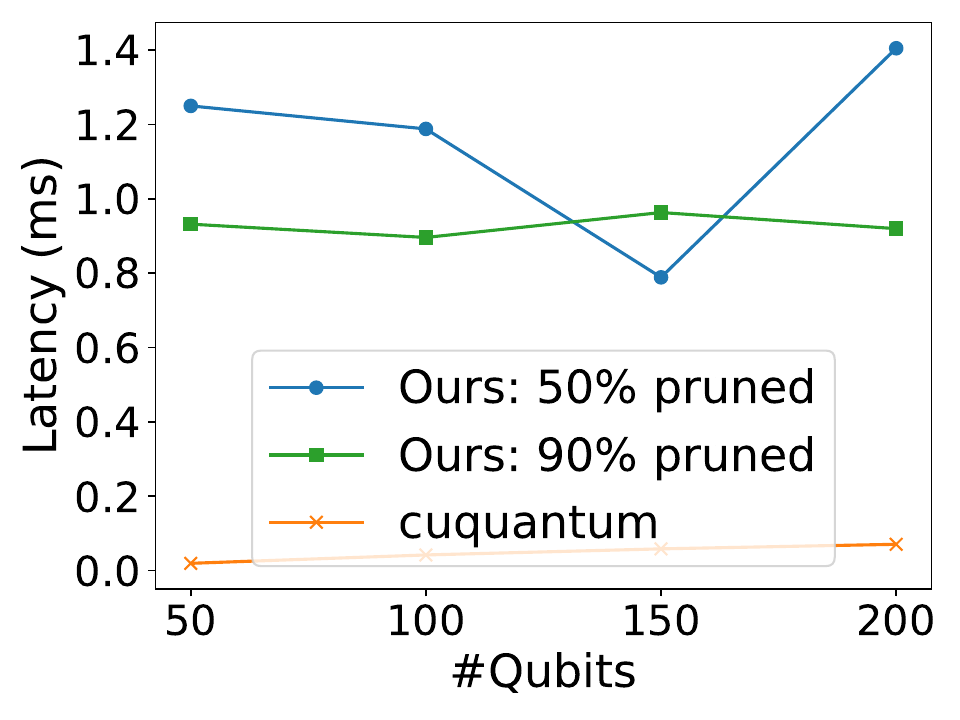}
        \caption{HWEA(\#Qubits,3,20) latency}
        \label{fig:HWEA(N,3,20)_latency}
    \end{subfigure}
    \hfill
    \begin{subfigure}{0.33\linewidth}
        \centering
        \includegraphics[width=\linewidth]{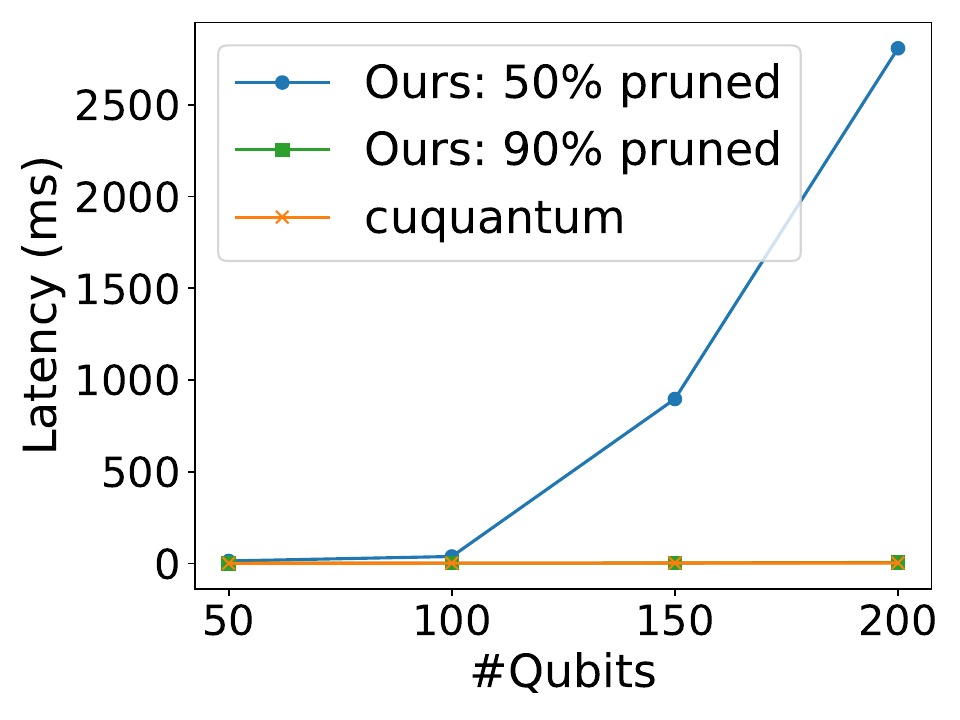}
        \caption{HWEA(\#Qubits,6,20) latency}
        \label{fig:HWEA(N,6,20)_latency}
    \end{subfigure}
    \begin{subfigure}{0.33\linewidth}
        \centering
        \includegraphics[width=\linewidth]{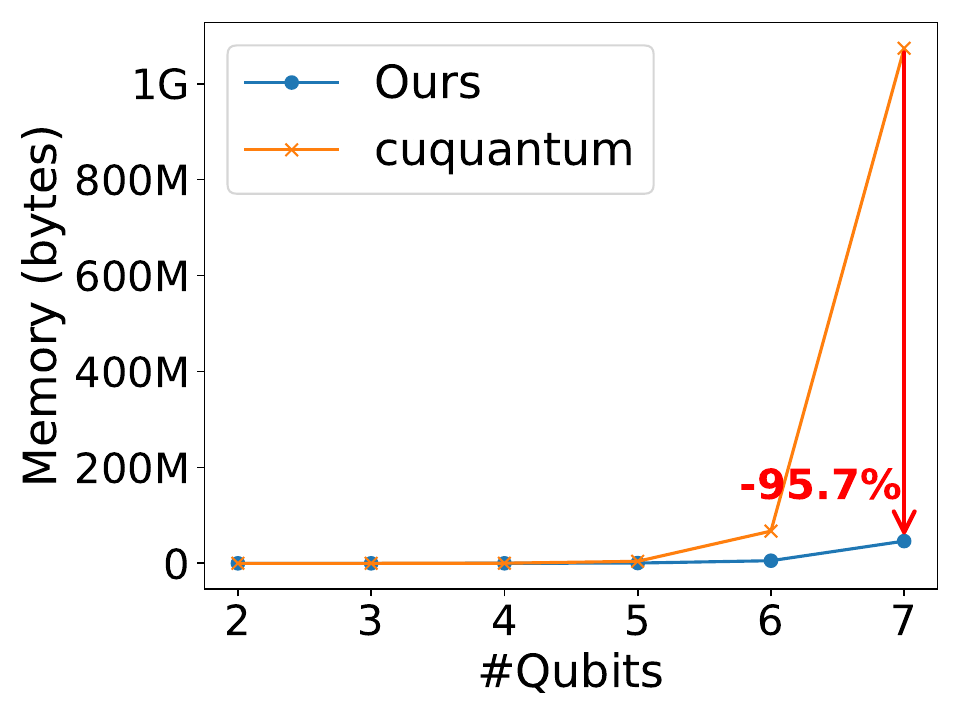}
        \caption{QFT memory footprint}
        \label{fig:qft_memory}
    \end{subfigure}
    \hfill
    \begin{subfigure}{0.33\linewidth}
        \centering
        \includegraphics[width=\linewidth]{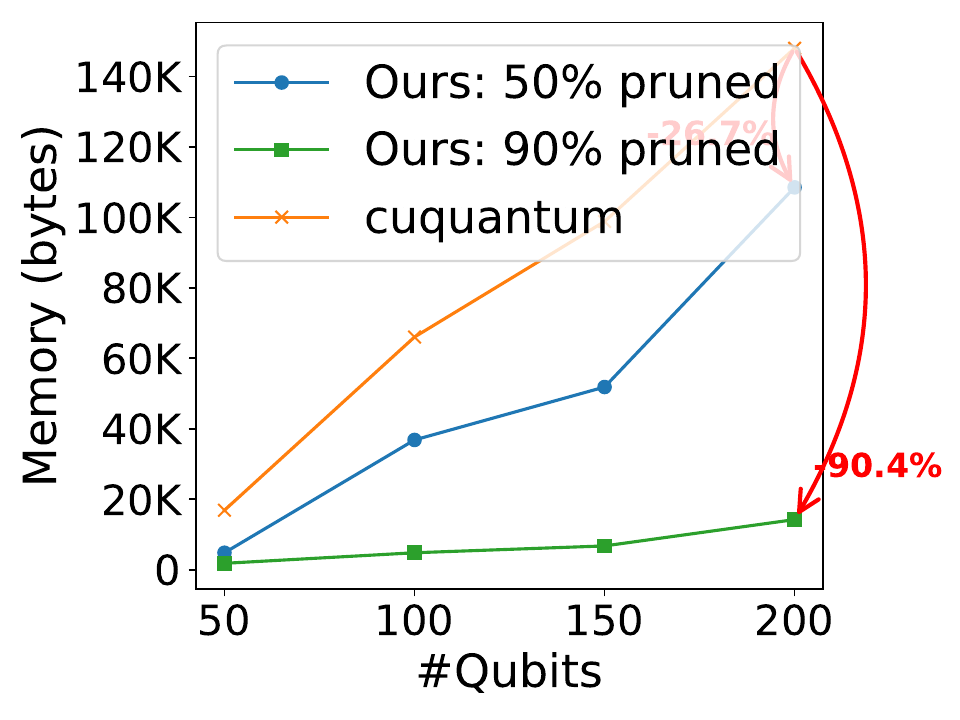}
        \caption{HWEA(\#Qubits,3,20) memory footprint}
        \label{fig:HWEA(N,3,20)_memory}
    \end{subfigure}
    \hfill
    \begin{subfigure}{0.33\linewidth}
        \centering
        \includegraphics[width=\linewidth]{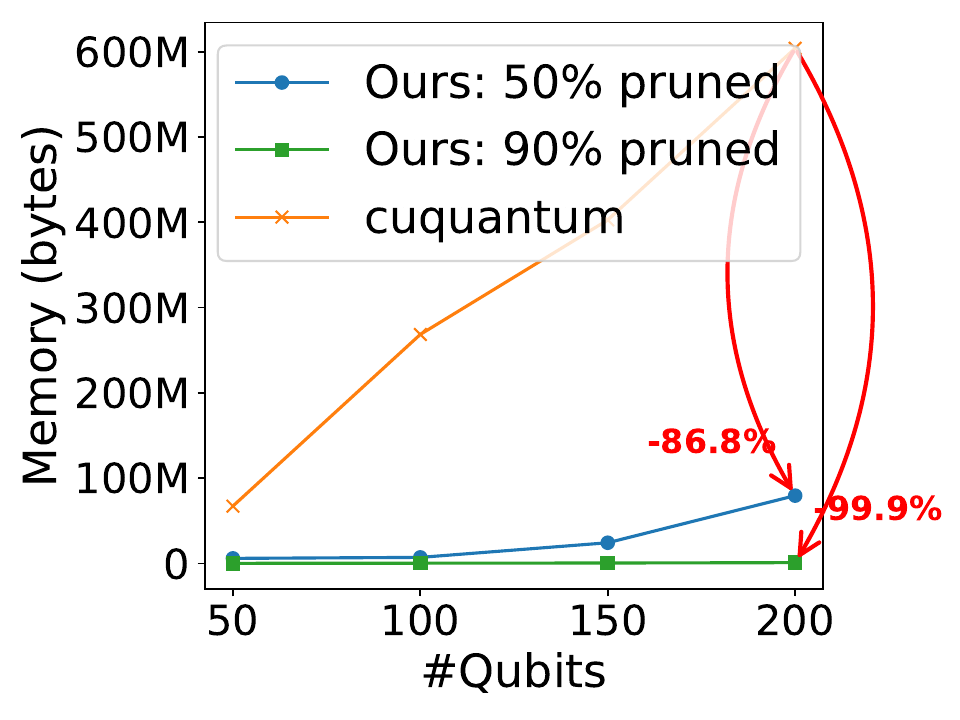}
        \caption{HWEA(\#Qubits,6,20) memory footprint}
        \label{fig:HWEA(N,6,20)_memory}
    \end{subfigure}

    \caption{
    The plot shows the postprocessing costs comparison between GPU cuQuantum implementation and our proposed sparse contraction approach.
    The x-axis is the number of qubits.
    The y-axis is the sparsity and performance metrics: FLOP count, latency, and memory footprint.
    The orange curves are cuQuantum.
    The blue and green curves in HWEA are our method when the pruning ratio is 50\% and 90\%.
    In QFT, there's no pruning so we use blue curves to represent our method.
    As the number of qubits grows, our method outperforms cuQuantum in FLOP count and memory footprint but is outperformed by cuQuantum in latency.
    In HWEA, the more pruning ratio our method has better performance.
    Our method's memory footprint reduction depends on sparsity.
    When sparsity is around 2\%, the reduction can be around 90\%.
    }
    \label{fig:postprocessing}
\end{figure*}

\subsection{Reduced Operations via Sparse Contraction}
\label{sec:flop_count}

As previewed in Section~\ref{sec:sparsity}, the factors and tensors in the circuit cutting task are sparse, and this is especially true for NISQ VQA circuits with parameter pruning.
Here we show that the sparsity has significant impact on the number of floating point operations needed for circuit cutting postprocessing, and this effect is orthogonal to the choice of tensor network contraction order.

\paragraph{NISQ versus beyond-NISQ circuit cutting workloads:}

We study the effect of sparsity on the feasibility of cutting two main types of quantum circuits.

For beyond-NISQ workloads, we select the quantum Fourier transform (QFT)\cite{jin2023quantumfouriertransformationcircuits} as a representative circuit.
These circuits are characteristically deep and have dense qubit interactions, leading to all-to-all connectivity.
Figure~\ref{fig:qft_sparsity} shows that even for these circuits, the subcircuit tensors become increasingly sparse.
Because QFT is by definition difficult to partition with few cuts, the QFT problem sizes that circuit cutting can handle are necessarily narrow with few qubits.
In the memory footprint limitations in our study, we are limited to reconstructing a seven-qubit QFT that requires 14 cuts.

For NISQ workloads, we again use the HWEA as a representative circuit that has uses in Hamiltonian simulation and machine learning.
As shown in Figure~\ref{fig:HWEA(N,3,20)_sparsity} and~\ref{fig:HWEA(N,6,20)_sparsity}, the sparsity of subcircuit tensors are controlled by the parameter pruning as expected, but also by the ansatz layer count, with deeper circuits having sparser tensors.
The HWEA circuit cutting workload can scale out to 200-qubits and six layers before reaching the memory limitations set in this study.

Additional workloads are studied in Table~\ref{tab:memory_footprint_comparison} to confirm the generality of our observation about sparsity.

\paragraph{Baseline versus our approach:}

{\footnotesize
\begin{algorithm}[t]
\begin{algorithmic}[1]
\caption{Sparse Tensor Contraction}
\label{alg:sparse_contract}
\Procedure{sparse\_contract}{sp\_tensor1, sp\_tensor2}
    \If {sp\_tensor1.size() $>$ sp\_tensor2.size()}
        \State swap(sp\_tensor1, sp\_tensor2)
    \EndIf
    \State \textbf{HashTable}$<$index, entry$>$ hash\_map
    \State \textbf{HashTable}$<$index, value$>$ result
    \State \textbf{List}$<$entry$>$ sp\_tensor\_result
    \State \# Preprocessing the index transformation
    \For{entry in sp\_tensor1}
        \State entry.common\_index$\gets$transform1(entry.index)
        \State entry.uncommon\_index$\gets$transform2(entry.index)
    \EndFor
    \For{entry in sp\_tensor2}
        \State entry.common\_index$\gets$transform3(entry.index)
        \State entry.uncommon\_index$\gets$transform4(entry.index)
    \EndFor
    \State \# Build the hash table
    \For{entry1 in sp\_tensor1}
        \State hash\_map[entry1.common\_index].append(entry1)
    \EndFor
    \State \# Contract
    \For{entry2 in sp\_tensor2}
        \State matched\_entries$\gets$hash\_map[entry2.common\_index]
        \For{entry1 in matched\_entries}
            \State new\_index$\gets$entry1.uncommon\_index\\+entry2.uncommon\_index
            \State result[new\_index]+=entry1.value*entry2.value
        \EndFor
    \EndFor
    \State sp\_tensor\_result$\gets$to\_list(result)
    \State \Return sp\_tensor\_result
\EndProcedure
\end{algorithmic}
\end{algorithm}
}



Our approach for circuit cutting postprocessing uses sparse tensor contraction, as described in Algorithm \ref{alg:sparse_contract}.
Instead of storing tensors in a flat format, sparse tensors use a list of index-value pairs, storing only non-zero values.
When contracting two sparse tensors, we build a hash table for the first tensor and process the second tensor by matching entries.
Our sparse algorithm is run on an AMD EPYC 7313 16-Core Processor.

For the baseline, we use Cotengra to plan a near-optimal contraction order~\cite{Gray2021hyperoptimized}, and contraction is done using cuQuantum~\cite{cuQuantum} running on a GeForce GTX 2080ti.
This represents the state-of-the-art backend for reconstruction.

\paragraph{Trends across ansatz gate pruning:}
As shown in Figures~\ref{fig:qft_flops},~\ref{fig:HWEA(N,3,20)_flops}, and~\ref{fig:HWEA(N,6,20)_flops}, our approach is able to perform the expectation value reconstruction in fewer floating point operations (FLOPs) avoiding zero operations due to sparsity.
The FLOP count of our proposed method would be around $\mathrm{(sparsity)}^2$ of that of cuQuantum.
The effect is more significant as the VQA ansatz parameters are pruned.

\paragraph{Trends across qubit count:}
Taking advantage of sparsity allows us to scale the HWEA workload to at least 200 qubits, which is at or beyond the qubit count circuit width attempted in previous work.
While prior work measured qubit count as a primary measure of whether circuit cutting can scale, the more nuanced viewpoint is that whether a circuit is easy to cut depends entirely on the type of circuit, as evidenced in the vastly different scalability of QFT vs. HWEA.

\paragraph{Trends across ansatz layer depth:}
The more significant controlling factor for whether circuit cutting can scale for NISQ VQAs is the layer count in the ansatzes.
The HWEA workload scales to six layers to match the order of magnitude number of FLOPs for the QFT workload.
Figures~\ref{fig:qft_latency},~\ref{fig:HWEA(N,3,20)_latency}, and~\ref{fig:HWEA(N,6,20)_latency} show that while our sparse approach incurs fewer FLOPs, the reduction in operations does not translate to improvements in wall-clock latency due to the parallelism offered in the GPU, while our hash-based approach requires pointer chasing.
However, these workloads are completed within seconds, and the primary limitation of workload scalability is actually the required memory footprint.

\begin{table*}[t]
    \centering
    \footnotesize
    \begin{tabular}{|c|c|c|c|c|c|c|}
        \hline
        \textbf{Workload} & \textbf{\#qubits} & \textbf{\#cuts} & \textbf{Sparsity} &  \begin{tabular}[c]{@{}c@{}}\textbf{cuQuantum}\\\textbf{memory footprint}\end{tabular}& \begin{tabular}[c]{@{}c@{}}\textbf{Our work's}\\\textbf{memory footprint}\end{tabular} &\begin{tabular}[c]{@{}c@{}}\textbf{Memory footprint}\\\textbf{reduction}\end{tabular} \\
        \hline
        UCCSD & 12 & 12 & 10.3\% & 128 MB & 26.5 MB & 79.3\% \\
        UCCSD & 14 & 14 & 0.56\% & 2 GB & 22.9 MB & 98.88\% \\
        QFT & 20 & 20 & 0.0009\%  & 8 TB & 190 MB & 99.998\% \\
        GHZ & 10 & 10 & 0.1\% & 8 MB & 17 KB & 99.8\%  \\
        GHZ & 20 & 20 & 0.00003\% & 8 TB & 5 MB & 99.999\% \\
        Erdos QAOA & 20 & 20 & 0.00005\% & 8 TB & 37.7 MB & 99.999\% \\
        Supremacy~\cite{huang2020classicalsimulationquantumsupremacy} & 16 & 16 & 1.0 \% & 34 GB & 714 MB & 97.9\% \\
        Sycamore~\cite{pednault2019leveragingsecondarystoragesimulate} & 16 & 16 & 0.001 \% & 34 GB & 1 MB & 99.997\% \\
        \hline
    \end{tabular}
    \caption{Memory footprint comparison between cuQuantum and our sparse approach for different circuit workloads.}
    \label{tab:memory_footprint_comparison}
\end{table*}

\begin{table*}[h!]
    \centering
    \footnotesize
    \begin{tabular}{|c|c|c|c|c|c|}
        \hline
        \textbf{Prior work}  & \textbf{Wire Cut} & \textbf{Gate Cut} & \textbf{NISQ workloads} & \textbf{Beyond-NISQ workloads} & \textbf{Open-sourced} \\ \hline
        CutQC~\cite{CutQC} & Yes & No  & \begin{tabular}[c]{@{}c@{}}HWEA (100 qubits),\\Supremacy (100 qubits),\\Approximate QFT (100 qubits)\end{tabular} & \begin{tabular}[c]{@{}c@{}} Grover (59 qubits),\\BV (100 qubits),\\Adder (100 qubits) \end{tabular} & \href{https://github.com/weiT1993/CutQC?tab=readme-ov-file}{Yes}\\ \hline
        FitCut~\cite{kan2024scalablecircuitcuttingscheduling} & Yes  & No & \begin{tabular}[c]{@{}c@{}}HWEA (100 qubits),\\Supremacy (72 qubits) \end{tabular} & \begin{tabular}[c]{@{}c@{}}BV(120 qubits),\\Adder(80 qubits)\end{tabular}  & Not yet  \\ \hline
        \begin{tabular}[c]{@{}c@{}}Integrated Qubit Reuse\\ and Circuit Cutting~\cite{pawar2023integratedqubitreusecircuit}\end{tabular} &  Yes  & Yes &  \begin{tabular}[c]{@{}c@{}} QAOA (50 qubits),\\ Supremacy (40 qubits) \end{tabular} & \begin{tabular}[c]{@{}c@{}} QFT (30 qubits),\\ Adder (40 qubits)\end{tabular} & Not yet  \\ \hline
        qiskit-addon-cutting~\cite{qiskit-addon-cutting} & Yes & Yes & N.A. & N.A. & \href{https://github.com/Qiskit/qiskit-addon-cutting}{Yes} \\ \hline
        This work & Yes & No & HWEA (200 qubits) & QFT (20 qubits) \textit{etc.} & Will be \\ \hline
    \end{tabular}
    \caption{Selected previous works that implemented and a circuit cutting framework.}
    \label{tab:previous_work}
\end{table*}

\subsection{Reduced Memory Footprint via Sparse Contraction}
\label{sec:memory_footprint}

Figures~\ref{fig:qft_memory},~\ref{fig:HWEA(N,3,20)_memory}, and~\ref{fig:HWEA(N,6,20)_memory} show the memory footprint reduction due to our sparse approach.
We can see that 2\% sparsity can reduce the memory footprint by around $\sim$90\% compared to cuQuantum.

The memory footprints are measured from the program implementations and checked against analytical models.
For our sparse implementation, we calculate the memory footprint of the two hash tables and three sparse tensor lists in Algorithm~\ref{alg:sparse_contract}.
We implement the function in C++ and we use vector and unordered\_map from the C++ standard template library as a list and hash table.
The memory footprints are modeled analytically as follows.
Let $A$ be the tensor size of a subcircuit1, $B$ be that of a subcircuit2, and $D$ that of the contracted tensor.
Without taking advantage of the sparsity, the memory footprint of cuQuantum is $4(A+B+D)$ bytes, where $4$ is the byte size of the floating point type.
Let the sparsity of the three tensors be $a,b,d$, respectively.
The memory footprint of our approach taking advantage of the sparsity is
{\small\[16(aA+bB+dD) + \text{hash\_table\_overhead}\times[\min(aA, bB)+dD]\]}

The first term describes the memory footprint of the three lists of entries, and the second term describes the memory footprint of the two hash tables.
The 16 in the first term is the size of the entry type.
In practice, the second term is negligible compared to the first term because the contraction path finder will always let us contract a small-sized tensor with a large-sized tensor.

Using the analytical model, we show additional experiments on memory footprint reduction for a benchmark of workloads shown in Table~\ref{tab:memory_footprint_comparison}.

\section{Related Work in Distributed QC}
\label{sec:distributed}


Distributed quantum computing is a set of techniques to combine the capabilities of multiple QCs~\cite{barral2024reviewdistributedquantumcomputing}.
Techniques include physically networking multiple QCs together so that qubits move between systems~\cite{10.1109/ISCA45697.2020.00051, laracuente2022modeling, Million_Qubit-Scale, 10.1145/3620665.3640377}, or entanglement between systems can be established before program runs begin~\cite{PhysRevLett.70.1895}.
A third strategy called circuit cutting is the use of classical computing to combine the results of multiple QCs, and is the focus of this paper.

Circuit cutting confers advantages and has several practical challenges.
The mathematics for cutting quantum circuits was proposed by Peng \textit{et al.}~\cite{Large_Small} and implemented for the first time by Tang \textit{et al.}~\cite{CutQC,tang2022scaleqcscalableframeworkhybrid}.
Numerous subsequent studies address different challenges in the circuit cutting strategy.
These include the decision on where to cut quantum circuits to fit the subcircuits into hardware constraints~\cite{kan2024scalablecircuitcuttingscheduling}.
The cut identification problem has been extended to consider both time-wise wire cuts and space-wise gate cuts~\cite{Gate_Cuts_and_Wire_Cuts,pawar2023integratedqubitreusecircuit}.
When the subcircuits are run, the resulting quantum state needs to be observed by efficient tomography~\cite{Golden_Circuit_Cutting_Points,Classical_Shadows}.
Finally, studies have been performed to reduce the reconstruction cost by approximation~\cite{Approximate_Reconstruction} and by identifying purely Clifford subcircuits~\cite{Clifford-based_Circuit_Cutting}.
Table~\ref{tab:previous_work} summarizes the most relevant prior work and the problem sizes they test.

\section{Conclusion}

This paper explores the influence of topology, determinism, and sparsity on quantum circuit cutting costs, with a focus on determinism and sparsity for subcircuit evaluation and reconstruction. Subcircuits are transformed into factor graphs, using knowledge compilation to identify minimal initialization and measurement sets based on non-zero input-output Pauli string pairs. This guides a more efficient tomography method, reducing zero-valued outputs and aiding error mitigation. Subcircuit factors are treated as tensors, and in reconstruction, tensors are contracted to exploit sparsity, reducing memory use. We compare our subcircuit execution to \textit{qiskit-addon-cutting} and classical shadows, and our reconstruction postprocessing to cuQuantum, showing clear advantages.

\bibliographystyle{plain}
\bibliography{references}

\end{document}